\documentclass[twocolumn,
               prd,
               aps,
               superscriptaddress,
               tightenlines,
               nofootinbib,
               eqsecnum,
               amsfonts,
               amsmath,
               longbibliography]{revtex4-1}

\usepackage{epsfig}
\usepackage{graphics}
\usepackage{graphicx}
\usepackage[dvipsnames,table]{xcolor}
\usepackage{bm}
\usepackage{natbib}
\usepackage{amssymb}
\usepackage{xspace}
\usepackage[normalem]{ulem} 
\usepackage[colorlinks]{hyperref}
\usepackage[caption=false]{subfig}
\usepackage{url}
\usepackage{float}
\usepackage[bottom]{footmisc}
\usepackage{lineno}
\usepackage{mathrsfs}
\usepackage{makecell}
\usepackage{microtype}
\usepackage{mathtools}
\usepackage{multirow}
\usepackage[inline]{enumitem}

\usepackage{acronym}
\def\newacronym#1#2#3{\gdef#1{\gdef#1{#2\xspace}#3 (#2)\xspace}}
\newacronym{\nr}{NR}{numerical relativity}
\newacronym{\eob}{EOB}{effective-one-body}
\newacronym{\gr}{GR}{general relativity}
\newacronym{\gw}{GW}{gravitational wave}
\newacronym{\snr}{SNR}{signal-to-noise ratio}
\newacronym{\qnm}{QNM}{quasinormal mode}
\newacronym{\pn}{PN}{post-Newtonian}
\newacronym{\imr}{IMR}{inspiral-merger-ringdown}
\newacronym{\rd}{RD}{ringdown}
\newacronym{\LVK}{LVK}{LIGO-Virgo-KAGRA}
\newacronym{\NQC}{NQC}{nonquasicircular}
\newacronym{\ROM}{ROM}{reduced order model}
\newacronym{\DINGO}{DINGO}{Deep Inference for Gravitational-wave Observations}

\AtBeginDocument{%
    \newwrite\bibnotes
    \def\bibnotesext{Notes.bib}
    \immediate\openout\bibnotes=\jobname\bibnotesext
    \immediate\write\bibnotes{@CONTROL{REVTEX41Control}}
    \immediate\write\bibnotes{@CONTROL{%
    apsrev41Control,author="08",editor="1",pages="1",title="0",year="1"}}
    \if@filesw
    \immediate\write\@auxout{\string\citation{apsrev41Control}}%
    \fi
}

\definecolor{aeired}{RGB}{173, 0, 0}
\definecolor{aeiblu}{RGB}{1, 67, 95}
\hypersetup{linkcolor=aeired}
\hypersetup{citecolor=aeiblu}
\hypersetup{urlcolor=aeiblu}

\usepackage{perpage}
\MakePerPage{footnote}

\newcommand{\lm}{{\ell m}}
\newcommand{\tm}{t^{\ell m}_{\rm match}}

\newcommand{\msun}{~{\rm M}_{\odot}}
\newcommand{\msunnt}{{\rm M}_{\odot}}

\newcommand{\pSEOB}{\texttt{pSEOBNRHM}}
\newcommand{\SEOB}{\texttt{SEOBNRHM}}
\newcommand{\hmpa}{\texttt{\detokenize{SEOBNRv4HM_PA}}}
\newcommand{\SEOBP}{\texttt{SEOBNRv4PHM}}
\newcommand{\SEOBE}{\texttt{SEOBNRv4EHM}}
\newcommand{\NRSur}{\texttt{NRSur7dq4}}
\newcommand{\IMRPHM}{\texttt{IMRPhenomXPHM}}

\newcommand{\pd}{\partial}
\newcommand{\dd}{{\rm d}}
\newcommand{\ii}{{\rm i}}

\newcommand{\AEI}{\affiliation{Max Planck Institute for Gravitational Physics (Albert Einstein Institute), D-14476 Potsdam, Germany}}
\newcommand{\UMD}{\affiliation{Department of Physics, University of Maryland, College Park, Maryland 20742, USA}}

\begin{document}

\title{Tests of general relativity in the nonlinear regime:
a parametrized \\ plunge-merger-ringdown gravitational waveform model}

\author{Elisa Maggio}        \AEI
\author{Hector O. Silva}     \AEI
\author{Alessandra Buonanno} \AEI \UMD
\author{Abhirup Ghosh}       \AEI

\date{\today}

\begin{abstract}
%
The plunge-merger stage of the binary-black-hole coalescence, when the bodies'
velocities reach a large fraction of the speed of light and the
gravitational-wave luminosity peaks, provides a unique opportunity to probe
gravity in the dynamical and nonlinear regime.
How much do the predictions of general relativity differ from the ones in other
theories of gravity for this stage of the binary evolution?
To address this question, we develop a parametrized waveform model, within the
effective-one-body formalism, that allows for deviations from general relativity
in the plunge-merger-ringdown stage. As first step, we focus on nonprecessing-spin,
quasicircular black hole binaries.
In comparison to previous works, for each gravitational wave mode, our model can modify, with respect to
general-relativistic predictions, the instant at which
the amplitude peaks, the instantaneous frequency at this
time instant, and the value of the peak amplitude.
We use this waveform model to explore several questions considering both
synthetic-data injections and two gravitational wave signals.
In particular, we find that deviations from the peak gravitational wave amplitude
and instantaneous frequency can be constrained to about 20\% with GW150914.
Alarmingly, we find that GW200129\textunderscore065458 shows a strong
violation of general relativity.
We interpret this result as a false violation, either due
to waveform systematics (mismodeling of spin precession) or due to data-quality
issues depending on one's interpretation of this event.
This illustrates the use of parametrized waveform models as tools to
investigate systematic errors in plain general relativity.
The results with GW200129\textunderscore065458 also vividly demonstrate the importance of waveform systematics and
of glitch mitigation procedures when interpreting tests of general relativity
with current gravitational wave observations.
\end{abstract}

\maketitle

\section{Introduction}
\label{sec:intro}

Remarkably, so far, the theory of \gr,
  introduced by Albert Einstein in 1915, has passed all available
  experimental and observational tests~\cite{Will:2014kxa}: on
  cosmological~\cite{Clifton:2011jh} and short
  scales~\cite{Kapner:2006si,Lee:2020zjt}, in the low-velocity,
  weak-field~\cite{GRAVITY:2020gka} and strong-field
  settings~\cite{Kramer:2021jcw,EventHorizonTelescope:2019pgp,EventHorizonTelescope:2022xqj}, and in the dynamical, high-velocity
  and strong-field
  regime~\cite{TheLIGOScientific:2016src,Abbott:2018lct,LIGOScientific:2019fpa,LIGOScientific:2020tif,LIGOScientific:2021sio}. The
  latter has been probed, since 2015, through the \gw observation of
  the coalescence of binary black holes (BBHs)~\cite{LIGOScientific:2016aoc,LIGOScientific:2016vlm,LIGOScientific:2018mvr,LIGOScientific:2020ibl,LIGOScientific:2021usb,LIGOScientific:2021djp}, neutron-star--black-hole (BH)
  binaries~\cite{LIGOScientific:2021qlt}, and binary neutron
  stars~\cite{TheLIGOScientific:2017qsa,LIGOScientific:2020aai} by the LIGO and Virgo detectors~\cite{LIGOScientific:2014pky,VIRGO:2014yos}.

Generally, tests of \gr{} with GW observations have been developed following two strategies:
theory independent and  theory specific. The former assumes that the underlying GW signal is
well-described by GR, and non-GR degrees of freedom (or parameters) are included to characterize any potential deviation.
These tests use GW observations to check consistency with their nominal predictions in GR, and
then constrain the non-GR parameters at a certain statistical level of confidence. Eventually, the
non-GR parameters can be translated to the ones in specific modified theories of gravity, albeit there could be
subtleties in doing it due to the choice of the priors and the actual parameters on which the measurements are done.
By contrast, analyses that compare directly the data with proposed modified theories of gravity belong to the
theory-specific framework of tests of GR.

Here, we focus on theory-independent tests of GR for BBHs. Historically, those tests have been
proposed introducing deviations in (or parametrizations of) the gravitational waveform, whether for the inspiral,
the merger or the ringdown stages, in time or frequency domain. Those parametrizations are clearly not unique;
neither they guarantee to fully represent the infinite space of modified gravity-theory waveforms. Furthermore,
non-GR parameters may be degenerate with each other, limiting the study to a subset of them~\cite{TheLIGOScientific:2016src}
 or demanding the use of principal-component-analysis methods~\cite{Saleem:2021nsb}.

Many parametrized waveforms have been suggested in the literature,
originally focusing on the inspiral phase~\cite{Blanchet:1994ez,Arun:2006yw,Arun:2006hn}, when the BBH system slowly
 but steadily looses energy through GW emission, and the bodies come closer and closer to each other until they merge.
When the first frequency-domain models for the \imr waveforms in GR became
available~\cite{Pan:2007nw,Ajith:2007kx}, a parametrized frequency-domain \imr
waveform model was proposed in Ref.~\cite{Yunes:2009ke}, variations of which were soon after employed in
Ref.~\cite{Sampson:2013jpa} for data-analysis explorations. Those initial works, together with other developments~\cite{Gossan:2011ha,Meidam:2014jpa}, are at the foundation of the
Test Infrastructure for GEneral Relativity (TIGER)~\cite{Li:2011cg,Agathos:2013upa,Meidam:2017dgf},
Flexible Theory Independent (FTI)~\cite{Mehta:2022pcn}, {\tt pSEOBNR}~\cite{Brito:2018rfr,Ghosh:2021mrv}, and {\tt pyRing}~\cite{Carullo:2018sfu,Carullo:2019flw,Isi:2019aib}  pipelines, which today are routinely used by the \LVK Collaboration~\cite{TheLIGOScientific:2016src,Abbott:2018lct,LIGOScientific:2019fpa,LIGOScientific:2020tif,LIGOScientific:2021sio} to perform parametrized tests of GR, probing the generation of GWs and the remnant properties, in the linear and nonlinear strong-field gravity regime.
Other theory-independent tests were also performed, e.g., in Refs.~\cite{Carullo:2018gah,Isi:2019asy,Tsang:2019zra,Bhagwat:2021kfa,Okounkova:2021xjv,Wang:2021gqm,Saleem:2021vph,Haegel:2022ymk}.

In this manuscript, we develop a parametrized time-domain \imr waveform model within the \eob{} formalism~\cite{Buonanno:1998gg,Buonanno:2000ef,Damour:2000we,Damour:2001tu,Buonanno:2005xu,Barausse:2009xi,Damour:2008gu,Pan:2010hz}. The \eob approach builds
semianalytical \imr waveforms by combining analytical predictions for the inspiral [notably from post-Newtonian (PN), post-Minkowskian (PM), and gravitational self-force (GSF) approximations] and ringdown phases (from BH perturbation theory) with physically-motivated Ans\"atze
for the plunge-merger stage. The \eob waveforms are then made highly accurate via a calibration to \nr waveforms of BBHs. The \eob formalism
relies on three key ingredients: the \eob conservative dynamics (i.e., a two-body Hamiltonian), the \eob radiation-reaction forces (i.e., the energy and angular momentum
fluxes) and the \eob{} \gw modes. Since the \eob waveforms are computed on the \eob dynamics by solving Hamilton's equations, in principle deviations from \gr can be introduced
in all the three building blocks, consistently. Here, for simplicity, following previous work~\cite{Brito:2018rfr,Ghosh:2021mrv} which focused on the ringdown stage, we
introduce non-GR parameters in the plunge-merger-ringdown \gw modes. We leave to future work the extension of the parametrization
to the conservative and dissipative dynamics, notably by including in the \eob dynamics fractional deviations to the PN (as well as PM and GSF) terms, to \nr-informed terms or specific
new terms motivated by phenomena observed in modified gravity theories.
We note that non-GR deviations in the EOB energy flux were implemented in Refs.~\cite{Ghosh:2016qgn,Ghosh:2017gfp}, and the corresponding EOB waveforms were used in IMR consistency and other tests of gravity in Refs.~\cite{Ghosh:2016qgn,Ghosh:2017gfp,Johnson-McDaniel:2021yge}.

Although the parametrized \imr model can in principle be constructed for precessing spinning BBHs, as first step, we consider nonprecessing BHs. There are two main
\eob families, {\tt SEOBNR} (e.g., see Refs.~\cite{Bohe:2016gbl,Cotesta:2018fcv,Ossokine:2020kjp}) and {\tt TEOBResumS} (e.g., see Refs.~\cite{Nagar:2018zoe,Nagar:2020pcj,Gamba:2021ydi}). We consider here the former, and in particular we focus on the \SEOB{} model developed in
Refs.~\cite{Bohe:2016gbl,Cotesta:2018fcv}, which contains \gw modes beyond the dominant quadrupole.
We denote the parametrized version \pSEOB. In Fig 1, we contrast a GR \SEOB{} waveform with parameters similar to the first GW observation, GW150914, with a \pSEOB \ waveform where the fractional deviations from GR are of the order of a few tens of percent. We can see that differences from \gr occur just
before, during, and after the merger stage, which is when the gravitational strain peaks.

\begin{figure*}[t]
\includegraphics[width=\textwidth]{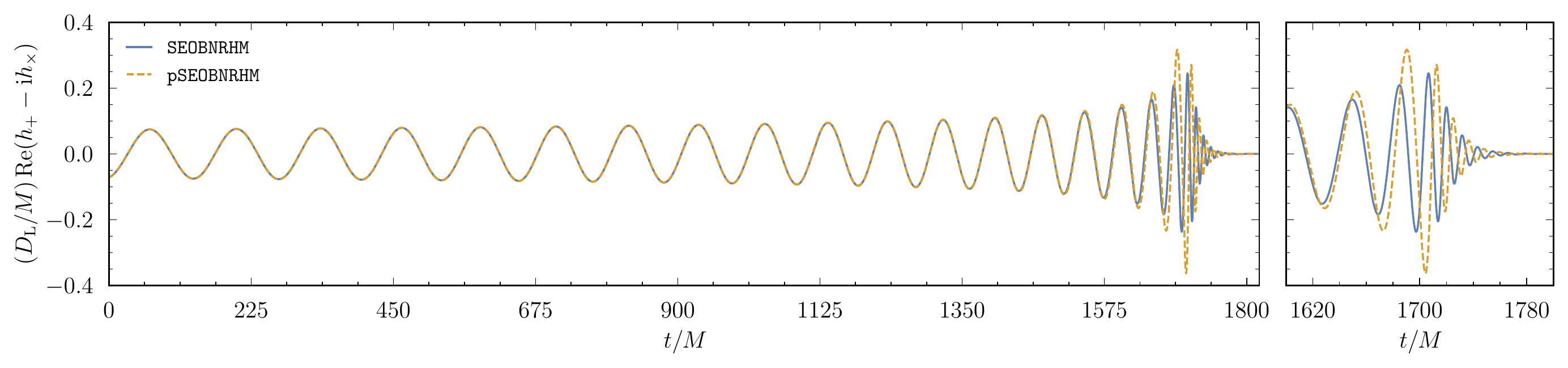}
\caption{Illustrative BBH waveform obtained with the \pSEOB{} model~introduced here (dashed line)
and the corresponding baseline model~\SEOB{}~\cite{Boyle:2008ge,Cotesta:2018fcv,Mihaylov:2021bpf}~(solid line) for a face-on, nonspinning and quasicircular
binary with GW150914-like mass-ratio $q = m_2 / m_1 \approx 0.867$,
and detector-frame total mass $M = m_1 + m_2 = 71.9 \msun$.
The \pSEOB~waveform is generated with non-\gr parameters values $\delta \Delta t = -0.2$, $\delta \omega = -0.4$,
and $\delta A = 0.5$.
These parameters change respectively, in comparison to \gr, the instant at
which the \gw amplitude peaks, the orbital frequency at this time instant, and
the value of the peak amplitude.
Both waveforms are phase aligned and time shifted around $20$~Hz using
the prescription of Refs.~\cite{Boyle:2008ge,Buonanno:2009qa,Pan:2010hz,Pan:2011gk}.
The details of how the waveform model is developed are in given Sec.~\ref{sec:waveform}, and additional details
about its morphology are presented in Sec.~\ref{sec:waveform_morpho}.
}
\label{fig:gw_ex}
\end{figure*}

The paper is organized as follows. In Sec.~\ref{sec:waveform}, we describe how we build the \pSEOB{} model starting
from the baseline model \SEOB{}, and introduce the non-GR parameters that describe potential deviations from GR during
the plunge-merger-ringdown stage. In Sec.~\ref{sec:waveform_morpho}, we study in detail the morphology of the parametrized
waveform, and understand which parts of the waveform change when the non-GR parameters are varied one at the time. After
discussing the basics of Bayesian analysis in Sec.~\ref{sec:pe}, we perform a synthetic-signal injection study in Sec.~\ref{sec:results_synth},
and then apply our parametrized \imr model to real data in Secs.~\ref{sec:gw150914} and \ref{sec:gw200129}, analyzing two events, GW150914 and GW200129.
Finally, we summarize our conclusions and future work in Sec.~\ref{sec:conclusions}.

Unless stated otherwise, we work in geometrical units in which $G = 1 = c$.

\section{The parametrized plunge-merger-ringdown waveform model}
\label{sec:waveform}

In this section we first review the \gr waveform model developed within the \eob formalism.
In Sec.~\ref{sec:waveform_built_param}, we explain how we deform this baseline
model by introducing deformations away from \gr in the plunge-merger-ringdown
phase.
%

\subsection{A brief review of the effective-one-body
gravitational waveform model}
\label{sec:waveform_built_review}

The \gw signal produced by a spinning, nonprecessing, and quasicircular BBH with component masses $m_{1}$ and $m_{2}$, and total mass $M = m_{1} + m_{2}$, is described in \gr by a set of eleven parameters, $\bm{\vartheta}_{\rm GR}$, given by
\begin{align}
    \bm{\vartheta}_{\rm GR} &= \left\{m_{1}, m_{2}, \chi_{1}, \chi_{2}, \iota, \psi,
    \alpha, \delta, D_{\rm L}, t_{c}, \phi_{c} \right\} \,,
    \label{eq:gr_params}
\end{align}
where $\chi_{i}$ ($i = 1, 2)$ are the constant-in-time projections of each BH's
spin vectors $\bm{S}_{i}$ in the direction of the unit vector
perpendicular to the orbital plane $\hat{\bm{L}}$, i.e.,
$\chi_{i} = \bm{S}_{i} \cdot \hat{\bm{L}} / m_{i}^{2}$, where $|\chi_{i}| \leqslant 1$,
$(\iota, \psi)$
describe the binary's orientation through the inclination and polarization angles,
$(\alpha, \delta)$ describe the sky location of the source in the detector frame,
$D_{\rm L}$ is the luminosity distance, and $t_{c}$ and $\phi_{c}$ are the reference time and phase, respectively.
It is convenient to define the chirp mass ${\cal M} = M \nu^{3/5}$, where $\nu = m_1 m_2 / M^2$
is the symmetric mass ratio, the asymmetric mass ratio $q = m_2 / m_1$, and
the effective spin $\chi_{\rm eff} = (\chi_1 m_1 + \chi_2 m_2) / M$.
We adopt the convention that $m_1 \geqslant m_2$ and thus $q \leqslant 1$.

The \gw polarizations can be written in the observer's frame as
\begin{align}
    h_{+}(\iota, \varphi_0; t) - \ii h_{\times}(\iota, \varphi_0; t) = \sum_{\ell=2}^{\infty} \sum_{m=-\ell}^{\ell} {}_{-2} Y_{\ell m}(\iota, \varphi_0) h_{\ell m}(t),
    \nonumber \\
\label{eq:hpc}
\end{align}
where $\varphi_0$ is the azimuthal direction of the observer, where, without loss of generality, we set $\varphi_0 = \phi_c$, and ${}_{-2} Y_{\ell m}$
are the $-2$ spin-weighted spherical harmonics~\cite{Newman:1966ub}, $\ell$
is the angular number and $|m| \leqslant \ell$ is the azimuthal number of each \gw mode, $h_{\ell m}$.

We follow Refs.~\cite{Ghosh:2021mrv,Silva:2022srr} and use as our baseline
model (i.e., the waveform model upon which the non-\gr deviation parameters are
added) the time-domain \imr waveform
developed in Refs.~\cite{Bohe:2016gbl,Cotesta:2018fcv,Mihaylov:2021bpf} within the \eob formalism~\cite{Buonanno:1998gg,Buonanno:2000ef,Damour:2000we,Damour:2001tu,Buonanno:2005xu,Barausse:2009xi,Damour:2008gu,Pan:2010hz}, \hmpa{}.\footnote{The model's name indicates that the \eob model ({\texttt{EOB}})
is calibrated to \nr simulations ({\texttt{NR}}), includes spin effects ({\texttt{S}}),
contains high-order radiation modes ({\texttt{HM}}), and uses the postadiabatic approximation ({\texttt{PA}})
to reduce the waveform generation time.
The version of the model used here is {\texttt{v4}}. The first version of this waveform family is the nonspinning {\texttt{EOBNRv1}} model of Refs.~\cite{Buonanno:2007pf,LIGOScientific:2011hqo}.}
The model uses the postadiabatic (PA) approximation, which was originally introduced in Refs.~\cite{Damour:2012ky,Nagar:2018gnk,Rettegno:2019tzh} (and also subsequently used in the \texttt{TEOBResumS} waveform models) to speed up the
generation of the time-domain waveforms for spinning, nonprecessing and quasicircular compact binaries. It includes the
$(\ell, |m|) = (2,2)$, $(2,1)$, $(3,3)$, $(4,4)$, and $(5,5)$ \gw modes.
For nonprecessing BBHs (i.e., with component spins
aligned or antialigned with the orbital angular momentum), we have that
$h_{\ell m} = (-1)^{\ell} \, h_{\ell-m}^{\ast}$.
Hence, we can consider $m > 0$ without loss of generality.
Hereafter, we refer to \hmpa{} as \SEOB{} for brevity.

As explained in Refs.~\cite{Bohe:2016gbl,Cotesta:2018fcv}, the \SEOB~waveform is constructed by attaching the merger-ringdown waveform, $h_{\ell m}^{\rm merger-RD}(t)$, to the inspiral-plunge waveform, $h_{\ell m}^{\rm insp-plunge}(t)$, at a matching time $t=\tm$,
\begin{align}
    \nonumber h_{\ell m}(t) &= h_{\ell m}^{\rm insp-plunge}(t) \, \Theta \left( t_{\rm match}^{\ell m}-t \right) \\
    &\quad + h_{\ell m}^{\rm merger-RD}(t) \, \Theta \left( t-t_{\rm match}^{\ell m} \right) \,,
\end{align}
where $\Theta(t)$ is the Heaviside step function and the value of $t_{\rm match}^{\ell m}$ is defined as
\begin{align}
t_{\rm match}^{\ell m} =
\begin{cases}
	t^{22}_{\rm peak} \,, &(\ell,m)=(2,2), \, (3,3),\, (2,1), \\
                         &\qquad\quad\,\,\,\,\, (4,4) \\
	t^{22}_{\rm peak} - 10M \,, &(\ell,m)=(5,5) \,,
\end{cases}
\label{eq:def_tmatch}
\end{align}
where $t^{22}_{\rm peak}$ is the time at which the amplitude of the $(2,2)$
mode [i.e., $h_{22}(t)$ in Eq.~\eqref{eq:hpc}] has its maximum value.
We impose that the amplitude and phase of $h_{\ell m}(t)$ at $t = \tm$
are $C^{1}$ (i.e., they are continuously and differentiable at this time instant).
The time $t^{22}_{\rm peak}$ is defined as
\begin{align}
	t^{22}_{\rm peak} = t^{\Omega}_{\rm peak} + \Delta t^{22}_{\rm peak} \,,
    \label{eq:t22_peak}
\end{align}
where $t^{\Omega}_{\rm peak}$ is the time in which the \eob orbital frequency peaks~\cite{Taracchini:2012ig}.
Calculations performed in the test-particle limit using BH perturbation theory found
that the amplitude and the orbital frequency peak at different
times, especially when the central BH has large spins~\cite{Damour:2007xr,Barausse:2011kb,Taracchini:2014zpa,Price:2016ywk}.
This motivates the introduction of the time-lag parameter $\Delta t^{22}_{\rm peak}$
in Eq.~\eqref{eq:t22_peak}, which can be fitted against \nr waveforms as function
of the symmetric mass ratio $\nu$ and the BH's spins $\chi_{1, 2}$ (see Sec.~II~B in Ref.~\cite{Bohe:2016gbl} for details).
We impose the condition $\Delta t^{22}_{\rm peak} \leqslant 0$ to
ensure that the attachment of the merger-ringdown waveform happens before the
peak of the orbital frequency, and thus before the end of the binary's dynamics.
For later convenience, we define
\begin{equation}
    \Delta t^{\rm GR}_{\ell m} = - \Delta t^{22}_{\rm peak}\,.
    \label{eq:Deltat_GR}
\end{equation}

Because we are interested in adding non-\gr terms to $h_{\ell m}^{\rm merger-RD}(t)$,
we now briefly review how the merger-ringdown waveform is constructed.
Further details can be found in Sec.~IV~E of Ref.~\cite{Cotesta:2018fcv}.
The merger-ringdown mode is written as
\begin{equation}
    h_{\lm}^{\rm merger-RD} = \nu \, \tilde{A}_{\lm}(t) \,
    e^{\ii \tilde{\phi}_\lm(t)} \,
    e^{\ii \sigma_{\lm 0} (t - t^{\lm}_{\rm match})} \,,
\label{eq:def_hmrd}
\end{equation}
where
$\sigma_{\lm 0}$ are the complex-valued
frequencies of the least damped \qnm of the remnant BH~\cite{Vishveshwara:1970zz,Press:1971wr,Detweiler:1980gk}.
We define $\sigma_{\lm 0}^{\rm R} = {\rm Im}(\sigma_{\lm 0}) < 0$
and $\sigma_{\lm 0}^{\rm I} = - {\rm Re}(\sigma_{\lm 0}) < 0$.
The functions $\tilde{A}_{\lm}$ and $\tilde{\phi}_\lm$ are given by~\cite{Bohe:2016gbl}
\begin{subequations}
\label{eq:merger_rd_funcs}
\begin{align}
    \tilde{A}_{\lm} &= c_{1,c}^{\lm} \,
    \tanh\left[ c_{1,f}^{\lm} \, (t - \tm) + c_{2,f}^{\lm} \right] + c_{2,c}^{\lm} \,,
    \label{eq:def_Alm}
    \\
    \tilde{\phi}_{\lm} &= \phi^{\lm}_{\rm match}
    - d_{1,c}^{\lm} \,
    \log \left[ \frac{1 + d_{2,f}^{\lm} \, e^{- d_{1,f}^{\lm} (t - \tm)}}{1 + d_{2,f}^{\lm}} \right] \,,
    \nonumber \\
    \label{eq:def_philm}
\end{align}
\end{subequations}
where $\phi^{\lm}_{\rm match}$ is the phase of the inspiral-plunge mode $h_{\lm}^{\rm insp-plunge}$ at
$t = t^{\lm}_{\rm match}$.
We see that Eqs.~\eqref{eq:merger_rd_funcs} depend on the set of parameters $c^{\lm}_i$ and $d^{\lm}_i$ ($i=1,2$),
which are either \emph{constrained} by imposing that $\tilde{A}_{\lm}$, $\tilde{\phi}_{\lm}$ are $C^{1}$ at
$t = \tm$ (we append the subscript ``$c$'') or \emph{free} parameters to be determined by
fitting against \nr waveforms (we append the subscript ``$f$'').

We now impose that $h_{\lm}$ is $C^{1}$ at $t = \tm$.
This yields two equations that relate the constrained coefficients
$c_{1,c}^{\lm}$ and $c_{2,c}^{\lm}$ to
the free coefficients $c_{1,f}^{\lm}$, $c_{2,f}^{\lm}$, to $\sigma_{\lm 0}^{\rm R}$ and to the
mode amplitude of $h_{\lm}^{\rm insp-plunge}$ and its first time derivative at the matching time, namely
$\vert h_{\lm}^{\rm insp-plunge}(\tm) \vert$
and $\pd_{t} \vert h_{\lm}^{\rm insp-plunge}(\tm) \vert$.
The equations are:
\begin{subequations}
\label{eq:cic}
\begin{align}
    c_{1,c}^{\lm} &= \frac{1}{\nu \, c_{1,f}^{\lm}}
    \left[ \pd_{t} \vert h_{\lm}^{\rm insp-plunge}(\tm) \vert \right.
    \nonumber \\
    &\left. \quad - \sigma_{\lm 0}^{\rm R} \, \vert h_{\lm}^{\rm insp-plunge}(\tm) \vert \right]
    \, \cosh^{2} c_{2,f}^{\lm}
    \,,
    \label{eq:c1c}
    \\
    c_{2,c}^{\lm} &= - \frac{1}{\nu} \, \vert h_{\lm}^{\rm insp-plunge}(\tm) \vert
    \nonumber \\
    &\quad + \frac{1}{\nu \, c_{1,f}^{\lm}} \left[ \pd_{t} \vert h_{\lm}^{\rm insp-plunge}(\tm) \vert \right.
    \nonumber \\
    &\left. \quad - \sigma_{\lm 0}^{\rm R} \, \vert h_{\lm}^{\rm insp-plunge}(\tm) \vert \right]
    \cosh c_{2,f}^{\lm} \, \sinh c_{2,f}^{\lm}\,.
    \nonumber \\
    \label{eq:c2c}
\end{align}
\end{subequations}
We also obtain one equation that relates the constrained parameter $d_{1,c}^{\lm}$
to the free coefficients $d_{1,f}^{\lm}$,
$d_{2,f}^{\lm}$, to $\sigma_{\lm 0}^{\rm I}$ and to the angular frequency of $h_{\lm}^{\rm insp-plunge}$ at the matching time. The latter is defined as $\omega_{\lm} = \dd \phi^{\rm insp-plunge}_{\lm} / \dd t$, where $\phi^{\rm insp-plunge}_{\lm} = {\rm arg}(h^{\rm insp-plunge}_{\lm})$ is the phase of the inspiral-plunge \gw mode.
The equation is,
\begin{equation}
    d_{1,c}^{\lm} = \left[ \omega_{\lm}^{\rm insp-plunge} (\tm) - \sigma_{\lm 0}^{\rm I} \right] \,
    \frac{1 + d_{2,f}^{\lm}}{d_{1,f}^{\lm} \ d_{2,f}^{\lm}}
    \,.
\label{eq:d1c}
\end{equation}
The values of
\begin{equation*}
\vert h_{\lm}^{\rm insp-plunge} \vert, \quad
\pd_{t} \vert h_{\lm}^{\rm insp-plunge} \vert, \quad {\rm and} \quad
\omega_{\lm}^{\rm insp-plunge} \,,
\end{equation*}
at $t = \tm$ are fixed by the so-called \NQC terms, $N_\lm (t)$.
The \NQC terms describe nonquasicircular corrections to the modes during the late inspiral and plunge.
The NQCs are a parametrized time series that is multiplied with the factorized PN GR modes, $h_{\lm}^{\rm F}$, such that the resultant time series is calibrated against NR simulations.
They are crucial in guaranteeing a very good agreement of the \SEOB{} amplitude and phase (relative to \nr) during the late inspiral and plunge.

The \gw modes in the inspiral-plunge part of the \eob waveform are given as
\begin{equation}
    h^{\rm insp-plunge}_{\lm}(t) = h_{\lm}^{\rm F}(t) \, N_{\lm}(t) \,,
\label{eq:def_iplg}
\end{equation}
where we refer the reader to Sec.~IV~C in Ref.~\cite{Cotesta:2018fcv} for details
on how $h_{\lm}^{\rm F}$ and $N_{\lm}$ are constructed.
For our purposes, it is sufficient to say that
$\vert h_{\lm}^{\rm insp-plunge} \vert(\tm)$,
$\pd_{t} \vert h_{\lm}^{\rm insp-plunge}(\tm) \vert$, and
$\omega_{\lm}^{\rm insp-plunge}(\tm)$ are the \emph{same} as the \nr values of
\begin{equation*}
\vert h_{\lm}^{\rm NR} \vert, \quad
\pd_{t} \vert h_{\lm}^{\rm NR} \vert, \quad {\rm and} \quad
\omega_{\lm}^{\rm NR}\,,
\end{equation*}
at $t = \tm$. The values of these three quantities are obtained for
each BBH, from the Simulating eXtreme Spacetimes (SXS) catalog of \nr waveforms~\cite{Boyle:2019kee},
after which a fitting formula that depends on the symmetric mass ratio $\nu$ and spins $\chi_1$ and $\chi_2$
is obtained to interpolate over the parameter space covered by the catalog.
Their explicit forms can be found in Ref.~\cite{Cotesta:2018fcv}, Appendix B.
At this point, we are left with the free parameters $c_{i,f}^{\lm}$ and $d_{i,f}^{\lm}$ ($i = 1, 2$) to fix.
This is accomplished through fits against \nr and Teukolsky equation-based
waveforms~\cite{Barausse:2011kb,Taracchini:2014zpa}, written also as functions of $\nu$,
$\chi_1$ and $\chi_2$.
The explicit form of these fits can be found in Ref.~\cite{Cotesta:2018fcv},
Appendix C.

\subsection{Construction of the parametrized model}
\label{sec:waveform_built_param}

With this framework established,
our strategy to develop a \emph{parametrized} \SEOB{}
\emph{model} (hereafter $\pSEOB$) is the following. We will introduce fractional deviations
to the \nr-informed formulas for the mode amplitudes and angular frequencies at $t = \tm$, i.e.,
\begin{subequations}
\label{eq:mod_h_and_omega}
\begin{align}
    |h_{\lm}^{\rm NR}| &\to |h_{\lm}^{\rm NR}| \, (1 + \delta A_{\lm})\,,
    \\
    \omega_{\lm}^{\rm NR} &\to \omega_{\lm}^{\rm NR} \, (1 + \delta \omega_{\lm})\,,
\end{align}
\end{subequations}
and we will also allow for changes to $\tm$ by modifying the time-lag parameter $\Delta t^{\rm GR}_{\ell m}$
[defined in Eq.~\eqref{eq:Deltat_GR}] as,
\begin{align}
\Delta t^{\rm GR}_{\ell m} \to \Delta t^{\rm GR}_{\ell m} \, \left( 1 + \delta \Delta t_{\ell m} \right) \,,
\label{eq:mod_dt}
\end{align}
where we constrain $\delta \Delta t_{\lm} > -1$ to ensure that $\tm$ remains less than
$t_{\rm peak}^{\Omega}$, and thus before the end of the dynamics, as originally
required~\cite{Boyle:2008ge,Cotesta:2018fcv}.
Equations~\eqref{eq:mod_h_and_omega} and~\eqref{eq:mod_dt} modify the constrained parameters $c_{i,c}^{\ell m}$
and $d_{i,c}^{\ell m}$ through Eqs.~\eqref{eq:cic}-\eqref{eq:d1c}, and consequently $\tilde{A}_{\ell m}$
and $\tilde{\phi}_{\ell m}$ that appear in the merger-ringdown waveform~\eqref{eq:def_hmrd} and are given by Eqs.~\eqref{eq:merger_rd_funcs}.
It is important to emphasize that Eqs.~\eqref{eq:mod_h_and_omega} and~\eqref{eq:mod_dt}
also modify the \NQC coefficients which enter the inspiral-plunge waveform in Eq.~\eqref{eq:def_iplg}.
This is because both $|h_{\ell m}^{\rm NR}|$ and $\omega_{\ell m}^{\rm NR}$ are
used to fix some parameters in the explicit form of $N_{\ell m}$. We refer the
reader to Refs.~\cite{Taracchini:2013rva,Bohe:2016gbl} and in particular to
Ref.~\cite{Cotesta:2018fcv}, Sec.~III~C, for details.
Hence, \emph{although we will refer to $\delta A_{\ell m}$, $\delta \omega_{\ell m}$, and $\delta \Delta t_{\ell m}$ as ``merger
parameters'' they, strictly speaking, also modify the plunge.}

We also introduce non-\gr deformations to the \qnm{s}, following the
same strategy applied in Refs.~\cite{Gossan:2011ha,Meidam:2014jpa,Brito:2018rfr,Isi:2019aib,Ghosh:2021mrv,Isi:2021iql}.
It consists in modifying the \qnm oscillation frequency and damping time,
defined respectively for the zero overtone $n=0$, as,
\begin{subequations}
\begin{align}
    f_{\lm 0} &= \frac{1}{2\pi} {\rm Re}(\sigma_{\lm 0}) = - \frac{1}{2\pi} \sigma^{\rm I}_{\lm 0} \,,
    \\
    \tau_{\lm 0} &= - \frac{1}{{\rm Im}(\sigma_{\lm 0})} = - \frac{1}{\sigma^{\rm R}_{\lm 0}} \,,
\label{eq:qnm_def}
\end{align}
\end{subequations}
according to the substitutions
\begin{subequations}
\begin{align}
    f_{\lm 0} &\to f_{\lm 0} \, (1 + \delta f_{\lm 0}) \,,
    \\
    \tau_{\lm 0} &\to \tau_{\lm 0} \, (1 + \delta \tau_{\lm 0}) \,, \label{tau}
\end{align}
\end{subequations}
and we impose that $\delta \tau_{\ell m 0} > -1$ to ensure that the remnant BH is stable
(i.e., it rings downs, instead of ``ringing-up'' exponentially). Note that in Refs.~\cite{Brito:2018rfr,Ghosh:2021mrv},
such deformations also concerned with the higher overtones, since the EOB model used for the merger-ringdown
included higher overtones.

Put it all together, we have the following set of plunge-merger-ringdown parameters:
\begin{align}
\bm{\vartheta}_{\rm nGR} &= \bm{\vartheta}_{\rm nGR}^{\rm merger} \, \cup \, \bm{\vartheta}_{\rm nGR}^{\rm RD}
    \label{eq:set_nonGR_params}
    \\
    &=
    \left\{
        \delta A_{\lm}, \delta \omega_{\lm}, \delta \Delta t_{\lm}
    \right\}
    \, \cup \,
    \left\{
        \delta  f_{\lm 0}, \delta \tau_{\lm 0}
    \right\} \,,
    \nonumber
\end{align}
intended to capture possible signatures of beyond-\gr physics in the most
dynamical and nonlinear stage of a BBH coalescence.
We will casually refer to them as ``non-\gr'' or as ``deformation'' (away from \gr) parameters.
In Table~\ref{tab:summary_of_nongr_params}, we summarize the $\bm{\vartheta}_{\rm nGR}$ parameters, their meaning, and the
constraints, if any, on their values.
The \gr limit is recovered when all parameters in $\bm{\vartheta}_{\rm nGR}$
are set to zero.

\begin{table}[t]
\begin{tabular}{c c | c c}
\hline
\hline
                          & Parameter                  &  Deformation  & Bound   \\
\hline
    \multirow{3}{*}{merger} & $\delta A_{\ell m}$   &  amplitude          &     \\
                          & $\delta \omega_{\ell m}$   &  instantaneous frequency         &         \\
                          & $\delta\Delta t_{\ell m}$  &  time lag         & $> -1$  \\
\hline
    \multirow{2}{*}{ringdown} & $\delta f_{\ell m 0}$      &  oscillation frequency         &          \\
                          & $\delta \tau_{\ell m0}$    &  damping time         & $>-1$         \\
\hline
\hline
\end{tabular}
\caption{
Summary of the non-GR parameters in the \pSEOB{} model.
The ringdown deformation parameters $\delta f_{\ell m 0}$ and $\delta
\tau_{\ell m 0}$ were introduced to the~\SEOB{} model in
Ref.~\cite{Ghosh:2021mrv}, while the merger deformation parameters
$\delta A_{\ell m}$, $\delta \omega_{\ell m}$, and $\delta \Delta t_{\ell m}$
are introduced here for the first time.
As explained in Sec.~\ref{sec:waveform_built_param}, although we call these
merger parameters, they do also affect the late inspiral-plunge part of the
waveform.
We quote under the column labeled ``bound'' the constraints on the parameter's values required by
our waveform model.
}
\label{tab:summary_of_nongr_params}
\end{table}

The \pSEOB{} model allows us to change the non-GR plunge-merger parameters $\bm{\vartheta}_{\rm nGR}^{\rm merger}$
for each ($\ell$, $m$) mode individually.
Here, for a first study, we will
assume that their values
are the same across different modes, that is to say,
\begin{equation}
\delta A_{\lm} = \delta A,\, \delta \omega_{\lm} = \delta\omega, \,\,
\textrm{and} \,\, \delta \Delta t_{\lm} = \delta \Delta t \,,
\end{equation}
for all the $\ell$ and $m$ modes in the waveform model.
This choice is motivated by the fact that
in GW150914 there are no significant changes in the posterior distributions of the binary parameters when using all the modes and only the $\ell=m=2$ mode.
As for the non-GR ringdown parameters $\bm{\vartheta}_{\rm nGR}^{\rm RD}$,
we will assume that they are nonzero only for the least-damped ($n=0$) $(2,2)$ mode.
Under these assumptions, we have a 16-dimensional parameter space to work with,
\begin{equation}\label{eq:params}
    \bm{\vartheta} = \bm{\vartheta}_{\rm GR} \, \cup \, \bm{\vartheta}_{\rm nGR}\,,
\end{equation}
where the \gr parameters $\bm{\vartheta}_{\rm GR}$ are defined in Eq.~\eqref{eq:gr_params}.

Some comments follow in order.
First, the parametrized deformation of \SEOB{} we have introduced is not unique. For instance, we could have added
additional fractional changes to $\partial_{t} |h_{\lm}^{\rm NR}|(\tm)$ or to the free parameters
in the merger-ringdown waveform segment [see Eq.~\eqref{eq:merger_rd_funcs}].
We have found a compromise between the number of new parameters
we can introduce and the physics we want to model; the optimal scenario being
that of having the most flexible \gw model that depends on the least number of deviation parameters.
In our case, we find the parameters ${\bm \vartheta}_{\rm nGR}$ defined in Eq.~\eqref{eq:set_nonGR_params} to be
sufficient for our purposes.
Second, one may fear that by effectively ``undoing'' the \nr calibration we would obtain nonphysical \gw{s}.
This is not the case, as shown in Fig.~\ref{fig:gw_ex} and as we will see
in Sec.~\ref{sec:waveform_morpho}.
Our model produces waveforms that are smooth deformations of the ones of \gr and
have sufficient flexibility to be applied
in tests of \gr (Secs.~\ref{sec:results_synth} and~\ref{sec:gw150914}) and
provide a diagnostic tool for the presence of systematic effects in \gr \gw models
(Sec.~\ref{sec:gw200129}).
%

\section{Waveform morphology}
\label{sec:waveform_morpho}

Having introduced our waveform model, we now discuss how each of the parameters
$\bm{\vartheta}_{\rm nGR}^{\rm merger}$ modify the \gw signal in \gr. An analogous exploration was done for $\bm{\vartheta}_{\rm nGR}^{\rm RD}$ in Ref.~\cite{Ghosh:2021mrv}, for this reason the present discussion is restricted to the merger parameters.
In each of the following sections, we vary the parameters $\delta A$, $\delta \omega$, and $\delta \Delta t$ one at a time.
We take the binary component masses and spins to be
\begin{equation}
    q = 0.867, \,
    \nu = 0.249,\,
    \chi_{1} = \chi_{2} = 0,\,
\label{eq:example_fixed_mass_spin}
\end{equation}
which are archetypal values of a GW150914-like event~\cite{TheLIGOScientific:2016wfe}, the inclination to be $\iota=0$
and, for clarity, we show
results only for $h_{22}$.
This is the dominant mode for such a quasicircular, nonspinning, and
comparable-mass BBH.
We end each section by showing how the waveform is modified when we apply
the deformations, with the same values, simultaneously to all \gw modes present
in \pSEOB{}.

\subsection{The amplitude parameter $\delta A$}

\begin{figure}[t]
\includegraphics[width=\columnwidth]{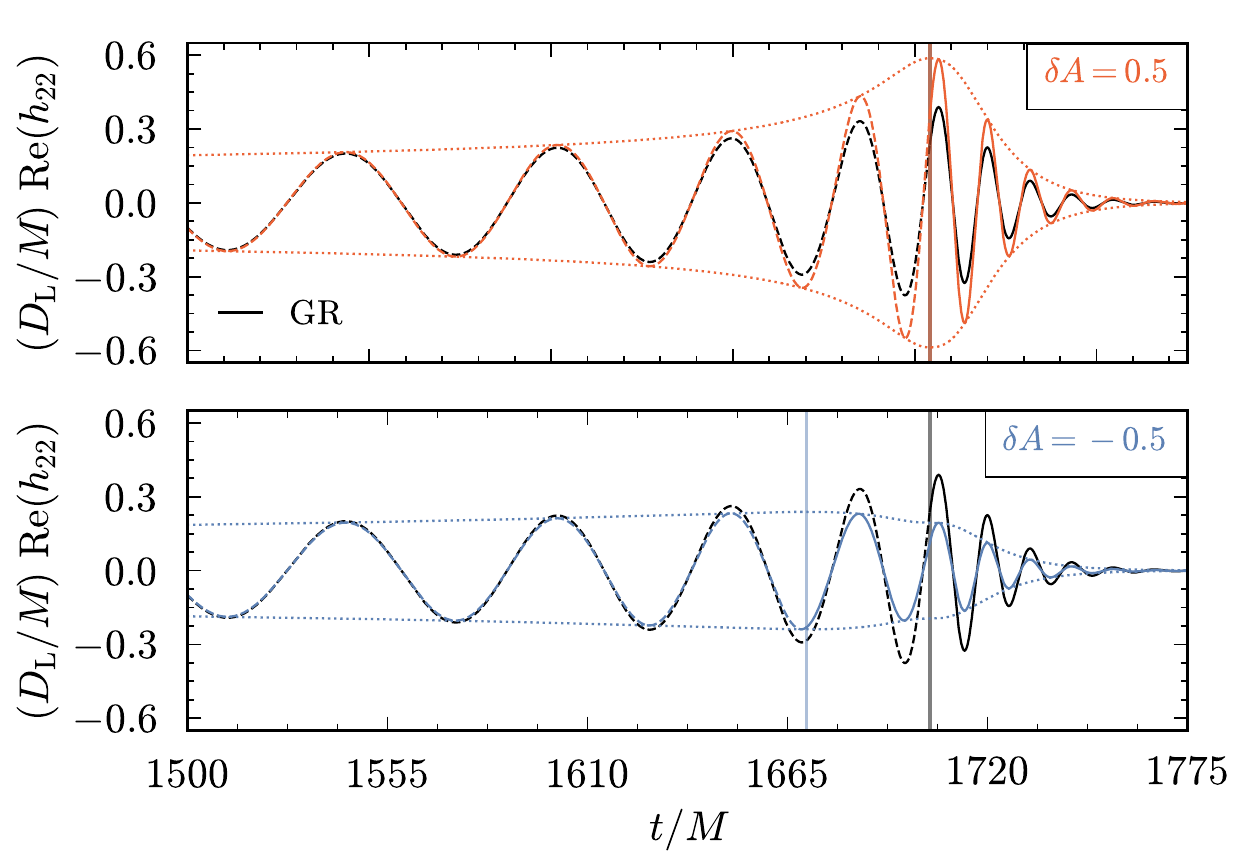}
\caption{
The time evolution near the merger of the real part of the $h_{22}$ mode for
nonzero values of the amplitude parameter $\delta A$.
We show the \gr prediction ($\delta A = 0$) with
the black lines.
Top panel: for $\delta A = 0.5$. Bottom panel: for $\delta A = -0.5$.
In both panels, we also show $\pm |h_{22}|$ for the non-\gr waveform (dotted lines), and
we use different line styles for the segment $t \leqslant t^{22}_{\rm match}$ (dashed lines)
and $t > t^{22}_{\rm match}$ (solid lines) for all waveform illustrated.
The matching times $t^{22}_{\rm match}$ are marked by the vertical lines.
}
\label{fig:h22_ex_dA}
\end{figure}

Let us start with $\delta A$, the amplitude parameter.
In Fig.~\ref{fig:h22_ex_dA} we show the real part
of $h_{22}(t)$, rescaled by the luminosity distance $D_{\rm L}$ and total mass
$M$, for two values of $\delta A$: $0.5$ (top panel) and $-0.5$ (bottom panel).
The dashed segment corresponds to $t \leqslant t^{22}_{\rm match}$ (i.e., the inspiral-plunge
part of waveform), whereas the solid segment corresponds to $t > t^{22}_{\rm match}$ (i.e.,
the merger-ringdown part of the waveform).
In both panels, the black
curve corresponds to the \gr signal ($\delta A$ = 0)
with the same binary parameters.
Both the \gr and non-\gr waveforms have been shifted in time and aligned in phase around $20$~Hz following the prescription
of Refs.~\cite{Boyle:2008ge,Buonanno:2009qa,Pan:2010hz,Pan:2011gk}.
The amplitudes of the non-\gr waveforms $\pm |h_{22}|$ are shown by the dotted
lines and form the envelope around ${\rm Re}(h_{22})$.

Unsurprisingly, for positive values of $\delta A$, the amplitude $|h_{22}|$ increases relative to its \gr value
while keeping $t^{22}_{\rm match} \approx 1704 \, M$ the same.
The situation is more interesting for $\delta A < 0$. For the binary under consideration, we find that
$|h_{22}|$ decreases for $\delta A \gtrsim -0.31$, but for $\delta A \lesssim -0.31$, we see that $\delta A$ pinches downwards the amplitude enough to result in
a local minimum (which we will refer to as $t^{22}_{\rm min}$) and two maxima, located before and after $t^{22}_{\rm min}$, with
the global maximum happening at $t^{22}_{\rm max} < t^{22}_{\rm min}$.
The values of both maxima are smaller than the \gr peak amplitude.
By construction, the matching time $t^{22}_{\rm match}$ is then shifted
to earlier times relative to its \gr value.
For the example of $\delta A =-0.5$ shown in the bottom panel of
Fig.~\ref{fig:h22_ex_dA}, the matching time is at approximately $1670 \, M$
(compare the location of the vertical lines in this panel).

\begin{figure}[t]
\includegraphics[width=\columnwidth]{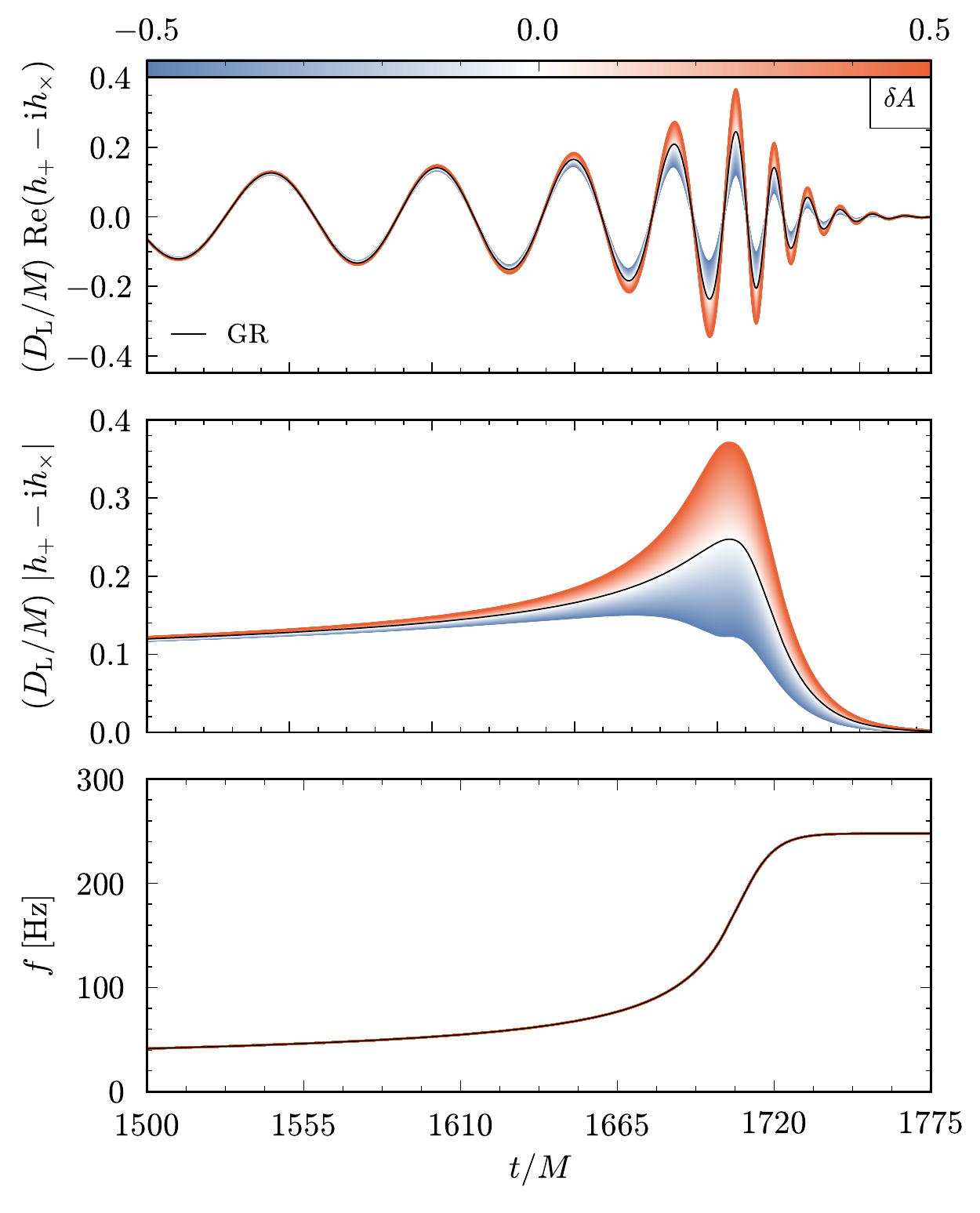}
\caption{The time evolution near the merger of the \gw strain for nonzero values of the
amplitude parameter $\delta A$, taken to affect in the same way all the $h_{\ell m}$ modes.
The \gr prediction ($\delta A = 0$) is shown by the black curves.
We show the real part of the strain (top panel), the strain amplitude (middle panel),
and the instantaneous frequency (bottom panel). As expected, the latter is unaffected
by the changes to the peak amplitude of the various \gw modes.}
\label{fig:gw_ex_dA}
\end{figure}

In Fig.~\ref{fig:gw_ex_dA}, we show a ``continuum'' of waveforms around the
time of merger, obtained by finely covering the interval $\delta A \in [-0.5, \,
0.5]$, and including $\delta A$ modifications to all modes in \pSEOB.
The \gr prediction is shown by the black solid line.
The top panel shows the real part of the strain, the middle panel the strain
amplitude, and the bottom panel the instantaneous frequency, defined as
$f = (2 \pi)^{-1} \, \dd \, {\rm arg}(h_{+} - \ii h_{\times}) / \dd t$.
As expected, we see that $f$ does not change by varying $\delta A$,
while the middle panel shows clearly how $\delta A$ changes the \gw amplitude. For negative values of $\delta A$, the presence of a local minimum in the \gw amplitude is evident, as discussed previously.

\subsection{The frequency parameter $\delta \omega$}

\begin{figure}[t]
\includegraphics[width=\columnwidth]{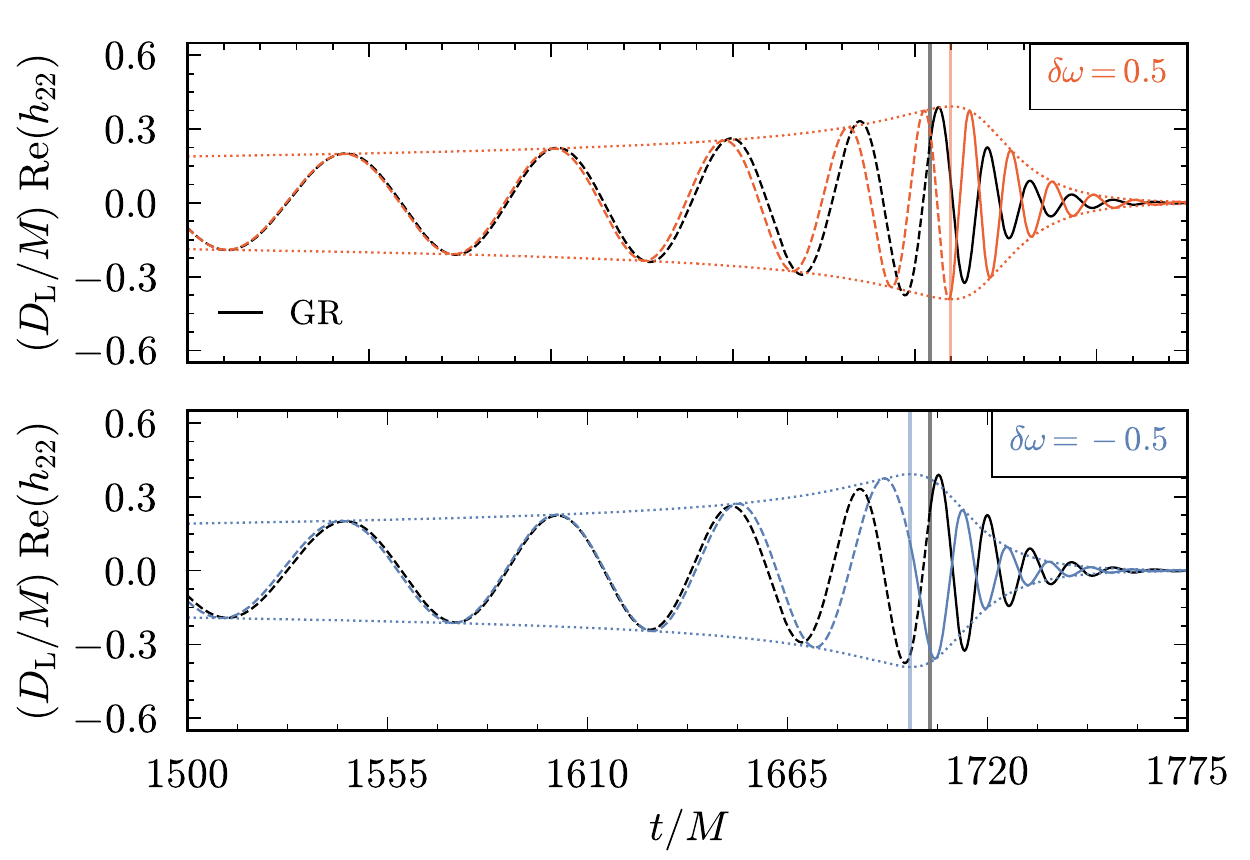}
\caption{
The time evolution near the merger of the real part of the $h_{22}$ mode for
nonzero values of the frequency parameter $\delta \omega$.
We show the \gr prediction
($\delta \omega = 0$) with the black lines.
Top panel: for $\delta \omega = 0.5$. Bottom panel: for $\delta \omega = -0.5$.
In both panels, we also show $\pm |h_{22}|$ for the non-\gr waveform (dotted lines), and
we use different line styles for the segment $t \leqslant t^{22}_{\rm match}$ (dashed lines)
and $t > t^{22}_{\rm match}$ (solid lines) for all waveform illustrated.
The matching times $t^{22}_{\rm match}$ are marked by the vertical lines.
}
\label{fig:h22_ex_dw}
\end{figure}

We now consider $\delta \omega$, the frequency parameter.
Figure~\ref{fig:h22_ex_dw} is analogous to Fig.~\ref{fig:h22_ex_dA},
except that we now consider $\delta \omega = 0.5$ (top panel) and $\delta \omega = -0.5$ (bottom panel).
We see that $\delta \omega$ induces a \emph{time-dependent} phase shift to the waveform, with its
effects being most noticeable near the merger, and causing $t_{\rm match}$ to happen later (earlier)
relative to \gr when $\delta \omega > 0$ ($\delta \omega < 0$), while keeping the peak amplitude
unaffected.

\begin{figure}[t]
\includegraphics[width=\columnwidth]{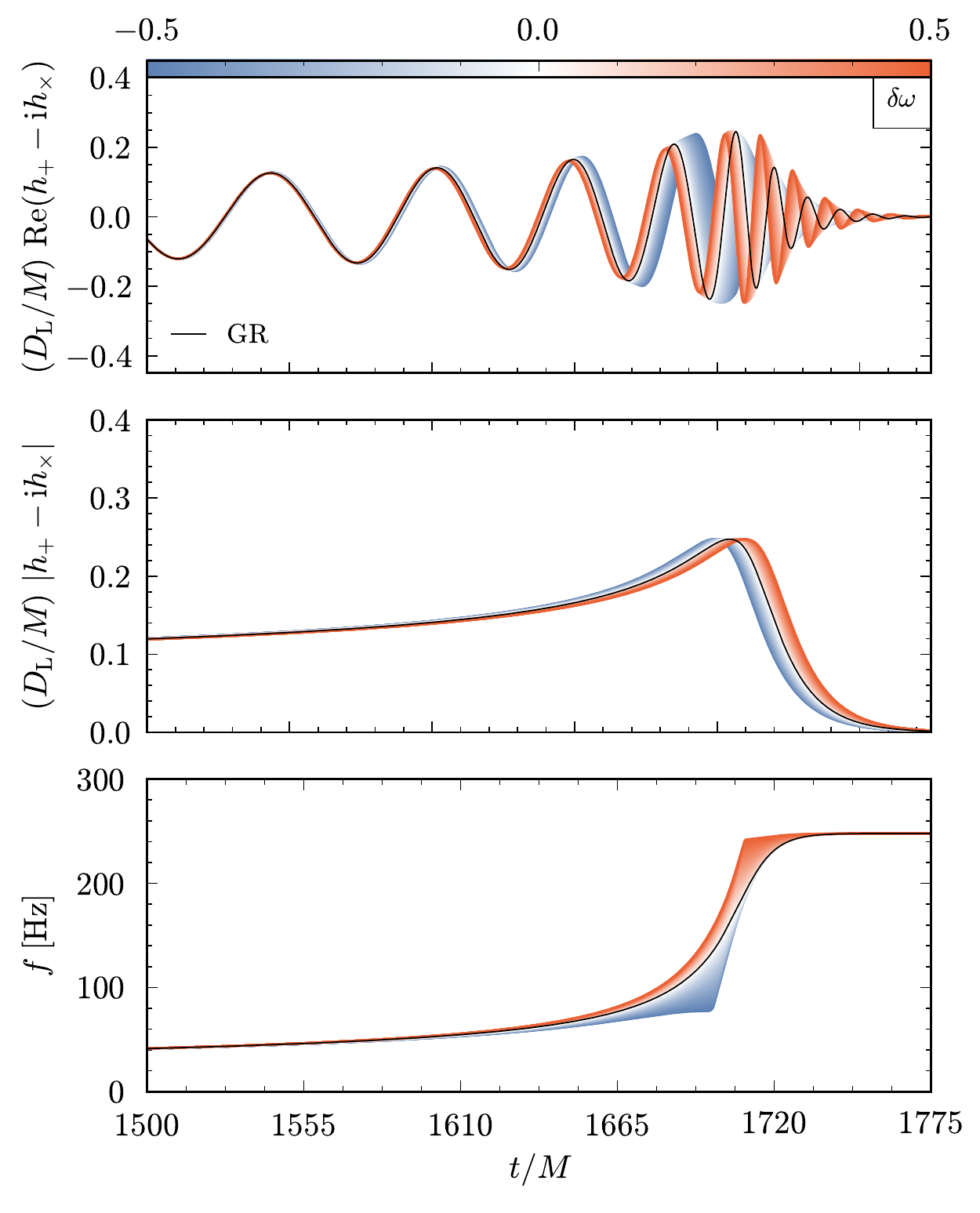}
\caption{The time evolution near the merger of the \gw strain for nonzero values of the
frequency parameter $\delta \omega$, assumed to be the same for all $h_{\ell m}$ modes.
The \gr prediction ($\delta \omega = 0$) is shown by the black curves.
We show the real part of the strain (top panel), the strain amplitude (middle panel),
and the instantaneous frequency (bottom panel).
In the top panel, we clearly see the phase difference between the non-\gr and \gr waveform near the merger.
This is partially due to the $\delta \omega$ itself, but also to the
phase-shift and time-alignment done with respect to the \gr waveform.
The effect of the latter is small as can be seen in the middle panel, which shows the amplitude.
The sharp changes to $f$ in the bottom panel for $|\delta \omega| \approx 0.5$
originate from us not imposing the continuity of the time derivative of $\omega_{\ell m}^{\rm NR}$ at $t = t_{\rm match}^{\ell m}$.
}
\label{fig:gw_ex_dw}
\end{figure}

In Fig.~\ref{fig:gw_ex_dw}, we show an analogous version of Fig.~\ref{fig:gw_ex_dA}, but now for $\delta \omega$.
Once more, the top panel shows the real part of the strain, the middle panel the strain amplitude,
and the bottom panel the instantaneous frequency. We focus on the region near the merger and
we plot the \gr curves ($\delta \omega = 0$) with black solid lines.
In the top panel, we can see the phase differences between the non-\gr and \gr waveforms, which are
the largest around the time of merger and ringdown.
This is in part due to the $\delta \omega$ itself, but also  to the
phase-shift and time-alignment procedure already mentioned, which we perform with
respect to the \gr waveform.
The effect of the latter is small, as can be seen in the middle panel for the amplitude,
where all curves nearly overlap in time.
In the bottom panel, we note sharp changes to $f$ when $|\delta \omega| \approx 0.5$.
They originate from us not imposing the continuity of the time derivative of
$\omega_{\ell m}^{\rm NR}$ at $t = t_{\rm match}^{\ell m}$~\cite{Bohe:2016gbl,Cotesta:2018fcv}.

\subsection{The time shift parameter $\delta \Delta t$}

\begin{figure}[t]
\includegraphics[width=\columnwidth]{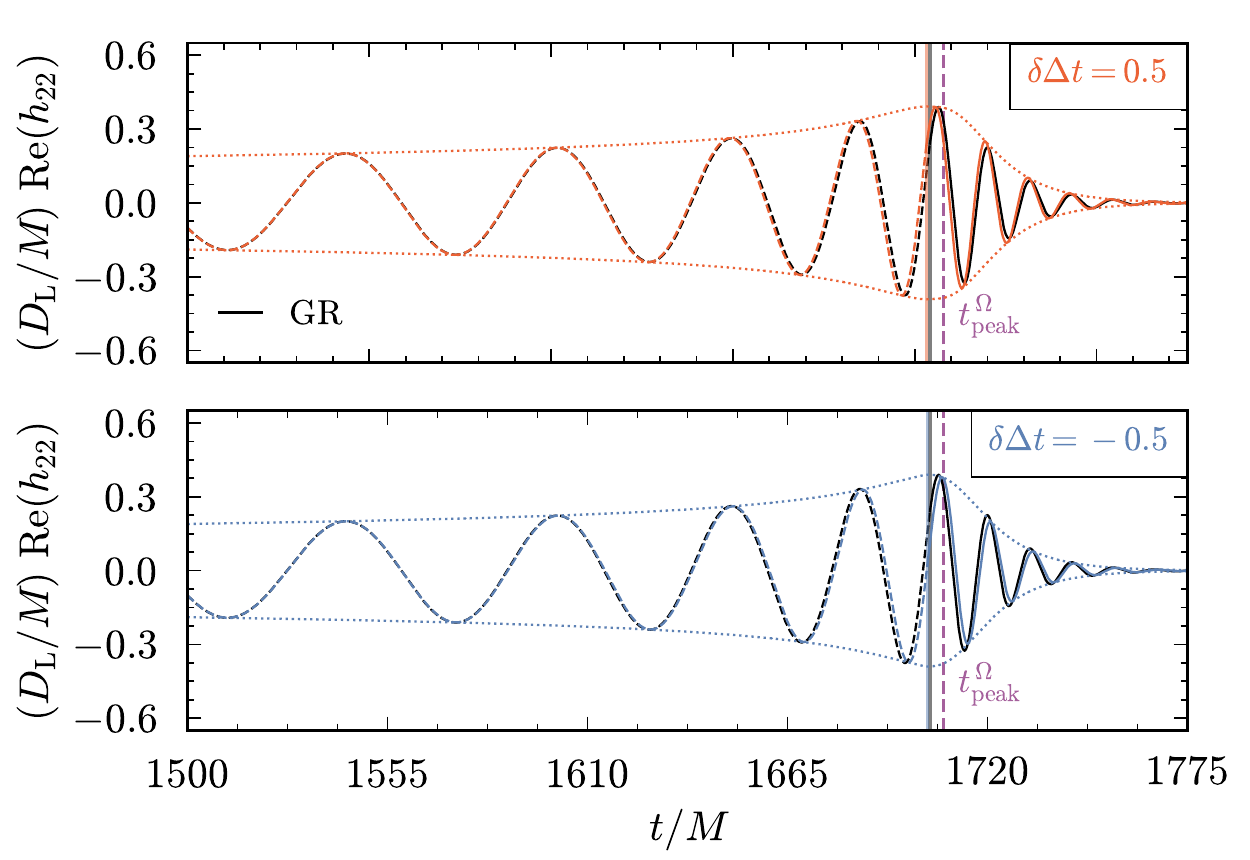}
\caption{
The time evolution near the merger of the real part of the $h_{22}$ mode for
nonzero values of the time shift parameter $\delta \Delta t$.
We show the \gr prediction
($\delta \Delta t = 0$) with the black lines.
Top panel: for $\delta \Delta t = 0.5$. Bottom panel: for $\delta \Delta t = -0.5$.
In both panels, we also show $\pm |h_{22}|$ for the non-\gr waveform (dotted lines), and
we use different line styles for the segment $t \leqslant t^{22}_{\rm match}$ (dashed lines)
and $t > t^{22}_{\rm match}$ (solid lines) for all waveform illustrated.
The matching times $t^{22}_{\rm match}$ are marked by the vertical lines.
For reference, we also show the instant in which the EOB frequency peaks ($t^{\Omega}_{\rm peak}$)
with vertical dashed lines.
}
\label{fig:h22_ex_dt}
\end{figure}

Finally, we now consider $\delta \Delta t$, the time shift parameter.
In Fig~\ref{fig:h22_ex_dt}, which is analogous to both Figs.~\ref{fig:h22_ex_dA} and~\ref{fig:h22_ex_dw},
we show waveforms for $\delta \Delta t = 0.5$ (top panel) and
$\delta \Delta t = -0.5$ (bottom panel).
Overall, we see small changes to the \gr waveform, in the form of an earlier $t_{\rm match}$ when $\delta \Delta t > 0$,
and later $t_{\rm match}$ when $\delta \Delta t < 0$.
Here, the changes due to the phase-shift and time-alignment are negligible, and the
shifts seen in the figure are due to $\delta \Delta t$.

\begin{figure}[t]
\includegraphics[width=\columnwidth]{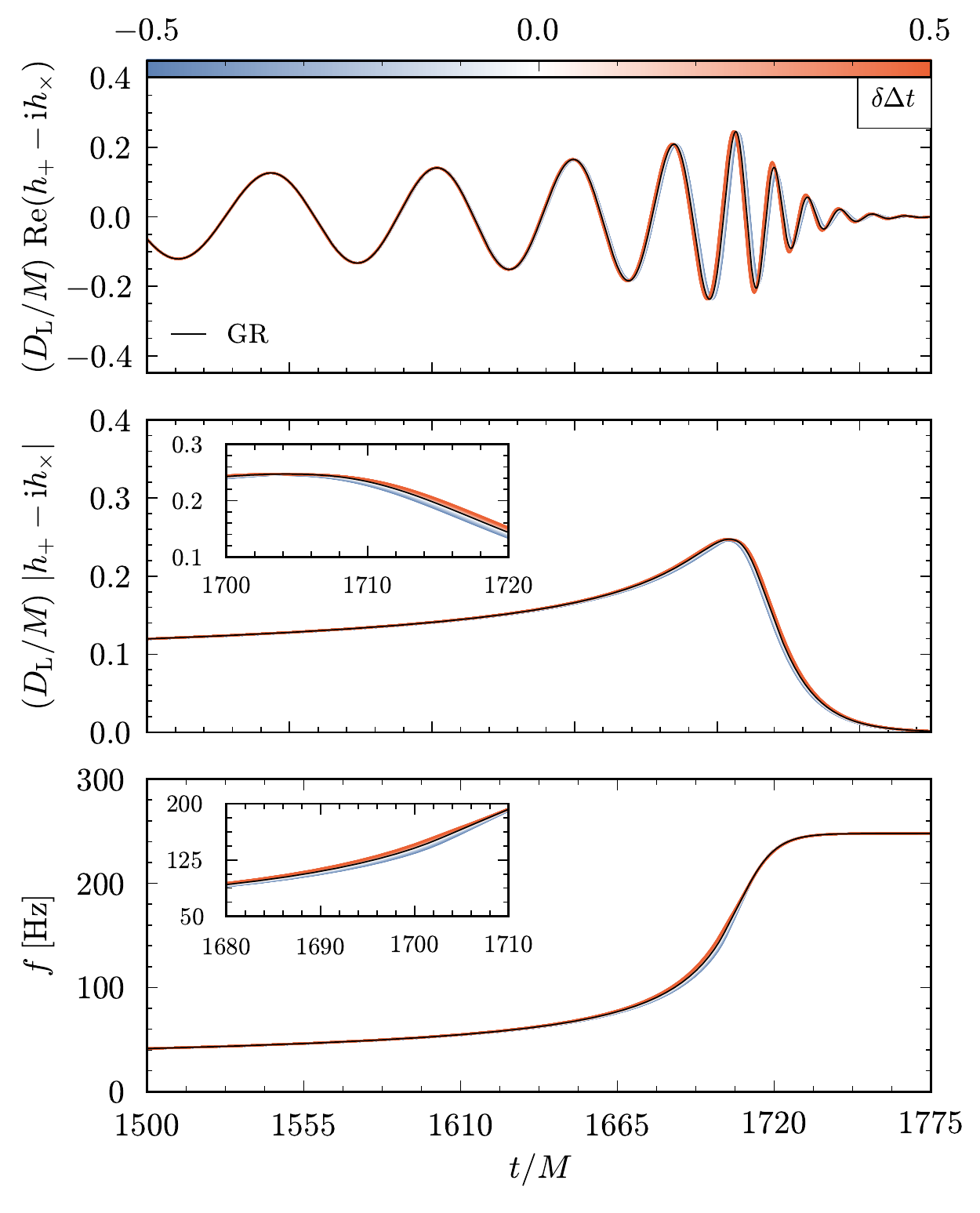}
\caption{The time evolution near the merger of the \gw strain for nonzero values of the
time shift parameter $\delta \Delta t$, assumed to be the same for all $h_{\ell m}$ modes.
The \gr prediction ($\delta \Delta t = 0$) is shown by the black curves.
We show the real part of the strain (top panel), the strain amplitude (middle panel),
and the instantaneous frequency (bottom panel).
The insets zoom into the time intervals $t / M \in [1700,\,1720]$ in the middle panel
and $t / M \in [1500,\,1775]$ in the bottom panel.
}
\label{fig:gw_ex_dt}
\end{figure}

Finally, in Fig.~\ref{fig:gw_ex_dt} we show a sequence of waveforms around
the time of merger, obtained by finely covering the interval $\delta \Delta t \in [-0.5, \, 0.5]$.
The \gr prediction is shown by the black solid line.
We see that the changes to the strain (top panel), its amplitude (middle panel),
and its frequency evolution (bottom panel) are small.
Therefore, $\delta \Delta t$ introduces changes to the \gr waveform which are in general
subdominant relative to those due to $\delta A$ and $\delta \omega$.
We also remark that $\Delta t^{\rm GR}_{\ell m}$ is not very sensitive to
the \eob calibration against \nr waveform.
Hence, the fractional changes we are introducing on $\Delta t^{\rm GR}_{\ell m}$
are comparable with the \nr fitting errors.
This explains why this parameter affects the \gr waveforms so little.

\section{Parameter estimation}
\label{sec:pe}

In the previous section, we have introduced our waveform model and discussed the
properties of the waveform morphology. Here, we summarize the Bayesian inference formalism
used for parameter estimation of \gw signals and synthetic-data studies. We describe the prior choices and the criteria for the \gw event selection.

\subsection{Bayesian parameter estimation}
\label{sec:pe_bayes}

Our hypothesis, $\mathcal{H}$, is that in the detector data, $d$, an observed \gw signal is
described by the waveform model \pSEOB. The model
\pSEOB \ has a set of \gr and non-\gr parameters, as in Eqs.~\eqref{eq:gr_params} and~\eqref{eq:set_nonGR_params}, where
\begin{equation} \label{param_non_GR}
\bm{\vartheta}_{\rm nGR} =
\left\{\delta A,\, \delta \omega,\, \delta \Delta t,\,
\delta f_{220},\, \delta \tau_{220}
\right\} \,.
\end{equation}
As said, we assume that the merger modifications are
the same for all $(\ell, m)$ modes present in the model \pSEOB{}.

The posterior probability distribution on the parameters of the model, $\bm{\vartheta}$, given the hypothesis, $\mathcal{H}$, is obtained using Bayes' theorem,
\begin{equation}\label{eq:bayes}
P(\bm{\vartheta}|d, \mathcal{H}) = \frac{P(\bm{\vartheta}|\mathcal{H})  P(d|\bm{\vartheta}, \mathcal{H}) }{P(d|\mathcal{H})} \,,
\end{equation}
where $P(\bm{\vartheta}|\mathcal{H})$ is the prior probability distribution, $P(d|\bm{\vartheta}, \mathcal{H})$ is the likelihood function, and $P(d|\mathcal{H})$ is the evidence of the hypothesis $\mathcal{H}$. For a detector with stationary, Gaussian noise and power spectral
density $S_{n}(f)$, the likelihood function can be written as
\begin{equation}
P(d|\bm{\vartheta}, \mathcal{H}) \propto \exp \left[ - \tfrac{1}{2} \left< d-h(\bm{\vartheta})|d-h(\bm{\vartheta}) \right>\right] \,,
\end{equation}
where the noise-weighted inner product is defined as
\begin{equation}
\langle A|B \rangle = 2\int_{f_{\rm low}}^{f_{\rm high}} \dd f \, \frac{\tilde{A}^*(f) \tilde{B}(f) + \tilde{A}(f) \tilde{B}^*(f)}{S_n(f)} \,,
\end{equation}
where $\tilde{A}(f)$ is the Fourier transform of $A(t)$, and the asterisk
denotes the complex conjugation, and $S_{n}(f)$ is the one-sided power spectral density of the detector.
The integration limits $f_{\rm low}$ and $f_{\rm high}$ set the bandwidth of the detector's sensitivity.
We follow the \LVK analysis and set $f_{\rm low} = 20~\text{Hz}$, while $f_{\rm
high}$ is the Nyquist frequency~\cite{LIGOScientific:2021djp}.
The posterior distributions are computed by using \texttt{LALInferenceMCMC}~\cite{Rover:2006ni,vanderSluys:2008qx},
a Markov-chain Monte Carlo that uses the Metropolis-Hastings algorithm to survey the likelihood surface and is implemented in
\texttt{LALInference}~\cite{Veitch:2014wba}, part of the \texttt{LALSuite} software suite~\cite{lalsuite}.

\subsection{Prior choices}
\label{sec:pe_priors}

The prior distributions on the GR parameters are assumed to be uniform in the component masses $(m_1, m_2)$, uniform and isotropic in the spin magnitudes $(\chi_1, \chi_2)$, isotropic on the binary orientation, and isotropically distributed on a sphere for the source location with $P(D_L) \propto D_L^2$.

For the non-GR parameters, as explained in Sec.~\ref{sec:waveform_built_param},
the internal consistency of the \pSEOB{} model requires that both $\delta
\Delta t$ and $\delta \tau_{220}$ are larger than $-1$
(cf.~Table~\ref{tab:summary_of_nongr_params}).
We use this fact to fix a common lower limit on the uniform priors on all
$\bm{\vartheta}_{\rm nGR}$. We set $1$ to be an upper limit on the uniform priors on the non-GR parameters.
This was sufficient in most of our analysis, but in a few cases we found that
the marginalized posteriors distributions for one or more non-GR parameters had support at $\bm{\vartheta}_{\rm nGR} \approx 1$.
In such cases we extended the priors' domains to $\bm{\vartheta}_{\rm nGR} \in
(-1,\,+2]$. Even at this wider range, we did not find anomalies in the waveform.
\subsection{Event selection}
\label{sec:pe_events}

The \pSEOB{} ringdown analysis performed in Ref.~\cite{LIGOScientific:2021sio} selected \gw events
from the GWTC-3 catalog~\cite{LIGOScientific:2021djp} which had a \snr~$\geqslant 8$ in the inspiral and post-inspiral regimes.
The requirement on the inspiral regime allows one to break the strong degeneracy between
the total mass of the binary and the ringdown deviation parameters~\cite{Brito:2018rfr,Ghosh:2021mrv}.
Among the \gw events that meet this criteria, two stand out in terms of their
constraining power on $\bm{\vartheta}_{\rm nGR}^{\rm RD}$, namely GW150914~\cite{LIGOScientific:2016aoc,LIGOScientific:2016vlm}
and GW200129\textunderscore065458 (hereafter GW200129)~\cite{LIGOScientific:2021djp}.
These two events, with a median total source-frame masses of $64.5\msun$ and $63.4\msun$,
respectively, are among the loudest BBH signals to date with a median total network \snr of
26.0 and 26.8, respectively~\cite{LIGOScientific:2021usb,LIGOScientific:2021djp}. GW150914 was detected by the two LIGO detectors at Hanford
and Livingston, whereas GW200129 was detected by the three-detector network of
LIGO Hanford, Livingston, and Virgo.

We guide ourselves by this result and use these two events to investigate what
constraints we can place on the merger-ringdown parameters.
We remark that this \snr selection criteria may be too strong if we are
interested in $\bm{\vartheta}_{\rm nGR}^{\rm merger}$ only.
We leave the study of the optimal \snr to constrain only the merger parameters
to a future work.

\section{Results: synthetic-signal injection studies}
\label{sec:results_synth}

\begin{flushleft}
\begin{table}[t]
\begin{tabular}{l | l}
\hline
\hline
Parameter (detector frame) & Value \\
\hline
Primary mass, $m_1$ [$\msunnt$] & 38.5 \\
Secondary mass, $m_2$ [$\msunnt$]  & 33.4 \\
Primary spin, $\chi_{1}$ & $3.47 \times 10^{-3}$ \\
Secondary spin, $\chi_{2}$  & $-4.40 \times 10^{-2}$ \\
Inclination, $\iota$ [rad]  & 2.69 \\
Polarization, $\psi$ [rad]  & 1.58 \\
Right ascension, $\alpha$ [rad]  & 1.22 \\
Declination, $\delta$ [rad]  & $-1.46$ \\
Luminosity distance, $D_{\text{L}}$ [Mpc]  & 337 \\
Reference time, $t_c$ [GPS]  & 1126285216 \\
Reference phase, $\phi_c$ [rad]  & 0.00 \\
\hline
\hline
\end{tabular}
\caption{
Values of the parameters $\bm{\vartheta}_{\rm GR}$ used in all synthetic-signal
injection studies in Sec.~\ref{sec:results_synth}.
The parameters are representative of GW150914, except for the luminosity
distance, which is chosen such that the total \snr, in a detector network
constituted by LIGO Hanford and Livingston operating at design sensitivity, is
approximately 100.
}
\label{tab:injected_values}
\end{table}
\end{flushleft}

In this section, we use \pSEOB{} to perform a number of synthetic-signal injection studies.
As we saw in Sec.~\ref{sec:waveform}, \pSEOB{} is a smooth deformation of the
\gr waveform model \SEOB{}, which is recovered when all $\bm{\vartheta}_{\rm nGR}$
parameters are set to zero.
This allows us to explore different scenarios that differ from one another on whether the
\gw signal and the \gw model used to infer the parameters of this signal
are described by \gr ($\bm{\vartheta}_{\rm nGR} = 0$) or not ($\bm{\vartheta}_{\rm nGR} \neq 0$).
We summarize these possibilities in Table~\ref{tab:injection_studies}.

To prepare the \gw signal we need to fix $\bm{\vartheta} =
\bm{\vartheta}_{\rm GR} \cup \bm{\vartheta}_{\rm nGR}$.
In all cases, we use values of $\bm{\vartheta}_{\rm GR}$ illustrative of a
GW150914-like BBH as in Table~\ref{tab:injected_values}.
We set all non-GR parameters to the same value, $\bm{\vartheta}_{\rm nGR} = 0.1$, whenever the
injected signal is non-GR.
By working exclusively with the \pSEOB{} waveform model, we avoid introducing
systematic errors due to waveform modeling in our analysis.
We also employ an averaged (zero-noise) realization of the noise to avoid
statistical errors due to noise.
The resulting \gw signal is then analyzed with the power spectral density $S_{n}(f)$ of
the LIGO Hanford and Livingston detectors both at design sensitivity~\cite{aLIGO_design_updated}.
In all cases, we set the distance to the binary to be such that the total network
\snr is approximately $100$.

In Sec.~\ref{sec:results_ngr_inj_gr_mod}, we do a preliminary
analysis where both injected and model waveforms are described by \gr.
This allows us to access the accuracy with which different binary parameters
can be recovered from the data in the detector network.
With these results as a benchmark, we can then proceed to inject a non-\gr
waveform and analyze it with a \gr model. This allows us to study the
systematic error introduced on the inferred binary parameters by assuming a
priori that \gr is true, while nature may not be so (the so-called fundamental
bias).
In Sec.~\ref{sec:results_gr_inj_ngr_mod}, we inject a \gr waveform and
try to recover its parameters with a non-\gr model. This allows us to
answer how much the non-\gr parameters can be constrained given an
event consistent with \gr.
Finally, in Sec.~\ref{sec:results_ngr_inj_ngr_mod}, we use non-\gr waveforms
as both our injection and our model. This answers whether we can
detect the presence of the non-\gr parameters in our signal.

\begin{table}[t]
\begin{tabular}{c c | c | c}
\hline
\hline
\ & \ & \multicolumn{2}{c}{Model}\\
\ & \ & GR & non-GR\\
\hline
\multirow{2}{*}{Injection} & GR & Sec.~\ref{sec:results_ngr_inj_gr_mod} & Sec.~\ref{sec:results_gr_inj_ngr_mod}\\
 & non-GR & Sec.~\ref{sec:results_ngr_inj_gr_mod} & Sec.~\ref{sec:results_ngr_inj_ngr_mod}  \\
\hline
\hline
\end{tabular}
\caption{
Summary of the synthetic-signal injection simulations performed in
Sec.~\ref{sec:results_synth}.
The label ``\gr" refers to the \SEOB~waveform model, whereas the label
``non-GR" refers to the \pSEOB~waveform model, where all
merger-ringdown parameters are set deviate in $10\%$ deviations relative
to their corresponding \gr values.
}
\label{tab:injection_studies}
\end{table}

\subsection{Fundamental biases on binary parameters}
\label{sec:results_ngr_inj_gr_mod}

We first explore the presence (or not) of biases in the inference of binary
parameters when the template waveform model assumes \gr, while the injected \gw
signal is non-GR~\cite{Yunes:2009ke,Vallisneri:2013rc}.
For this purpose, we first inject a synthetic \gr \gw signal with $\text{SNR} = 98$ and recover
the binary parameters with a \gr model. By doing this exercise first, we gain
an idea on the accuracy with which the parameters of the binary
(cf.~Table~\ref{tab:injected_values}) can be recovered in our set up.
Next, we repeat the same analysis but now using as our synthetic \gw signal
the one obtained with \pSEOB{}.
The signal is prepared using the same binary parameters $\bm{\vartheta}_{\rm
GR}$ shown in Table~\ref{tab:injected_values} with $\text{SNR} = 104$, but now we let
$\bm{\vartheta}_{\rm nGR} = 0.1$.

\begin{figure}[t]
\centering
\includegraphics[width=\columnwidth]{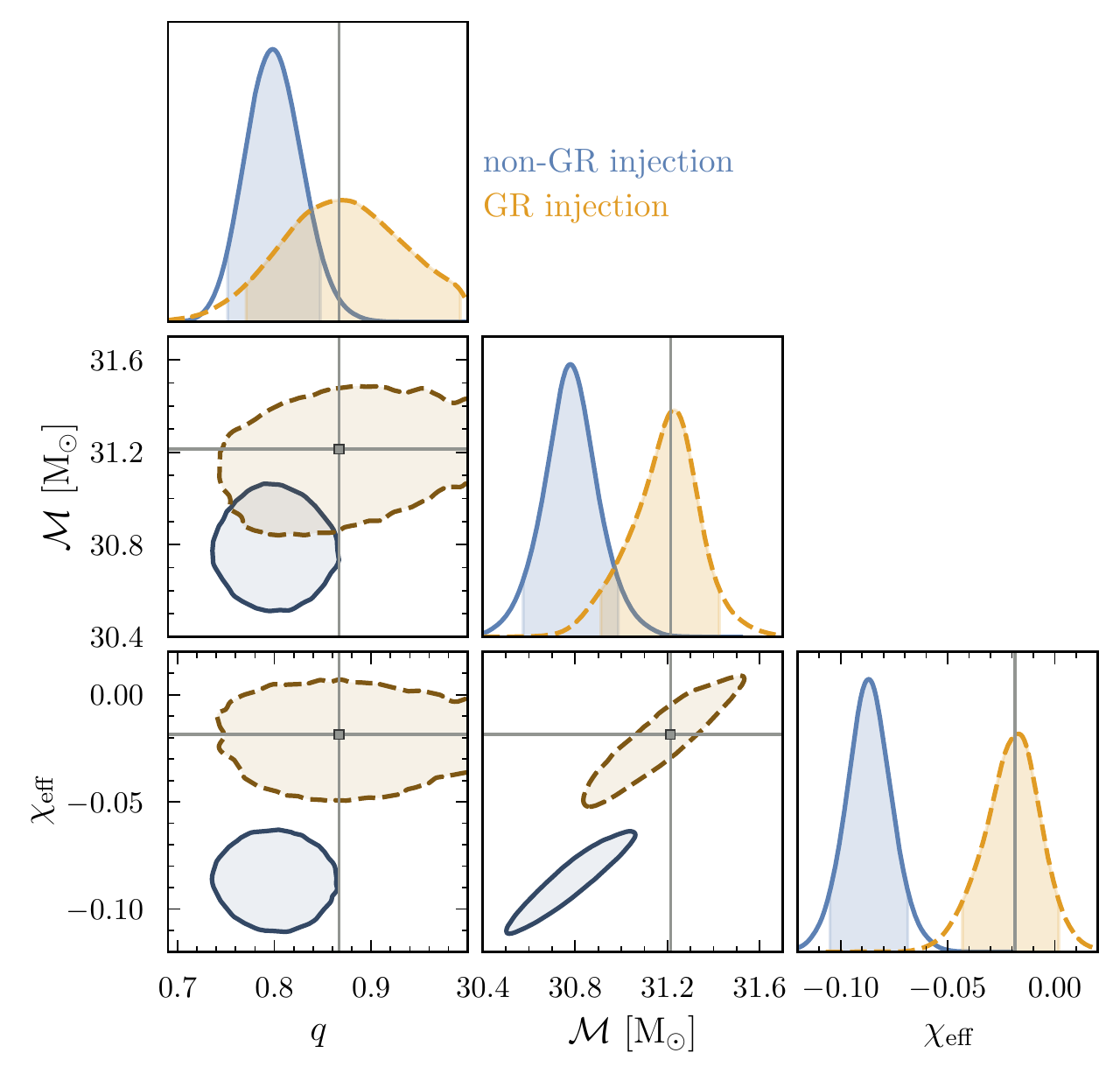}
\caption{
The one- and two-dimensional posterior distributions on the
intrinsic binary parameters of mass ratio $q$, detector-frame chirp mass
$\mathcal{M}$ and effective spin $\chi_{\rm eff}$ for a \gr injection (dashed
curve) and a non-\gr injection (solid curve) with $10\%$ deviations in the
merger-ringdown parameters $\bm{\vartheta}_{\rm nGR}$. All contours indicate
90\% credible regions.
The parameter estimation is performed assuming the GR \SEOB~waveform model. The
vertical and horizontal lines mark the injected values. The measurements
with non-\gr injections are visibly biased, most preeminently in $\chi_{\rm eff}$ and $\mathcal{M}$.
}
\label{fig:cornerplot_nonGRinj_GRrec}
\end{figure}

The results of our two analyses are shown in Fig.~\ref{fig:cornerplot_nonGRinj_GRrec}.
We show the one- and two-dimensional posterior distributions of a subset of the
intrinsic binary parameters, namely, the mass ratio $q$, the detector-frame chirp mass
$\mathcal{M}$ and the effective spin $\chi_{\rm eff}$.
In all panels, the ``true'' (injection) values of these parameters are marked
by the vertical and horizon lines.
We see that in the case of a non-\gr injection (solid curves), the posterior
distributions of the parameters are shifted from the injected values and from the
posterior distributions in the case of a \gr injection (dashed curves).
We attribute the differences in the 90\% contours of the posterior distributions
to the fact that in the non-GR injection a smaller value of the chirp mass
${\cal M}$ is inferred. This suggests that the GR waveform that best fits the
data has a longer inspiral and this makes the inference of the other binary
parameters more precise.
The recovered SNR from the GR analysis of the non-GR signal is almost the same as the injected one, i.e., $\text{SNR}=104$.
Hence, if a \gw signal with deviations from \gr would be analyzed by current
\gr templates, the \gw event would be interpreted as a BBH in \gr with
different values of the binary parameters.

\subsection{Constraints on deviations to general relativity}
\label{sec:results_gr_inj_ngr_mod}

We now inject a synthetic \gw signal in \gr using the parameters
$\bm{\vartheta}_{\rm GR}$ in Table~\ref{tab:injected_values} with $\text{SNR} = 98$.
We analyze the signal using the \pSEOB~waveform model, allowing both $\bm{\vartheta}_{\rm GR}$
in Eq.~\eqref{eq:gr_params} and $\bm{\vartheta}_{\rm nGR}$ in Eq.~\eqref{param_non_GR} to vary.
This simulates a scenario where we have a \gw event consistent with \gr and we want
to understand which constraints it places on the non-\gr parameters in our waveform model.

\begin{figure}[t]
\centering
\includegraphics[width=1.01\columnwidth]{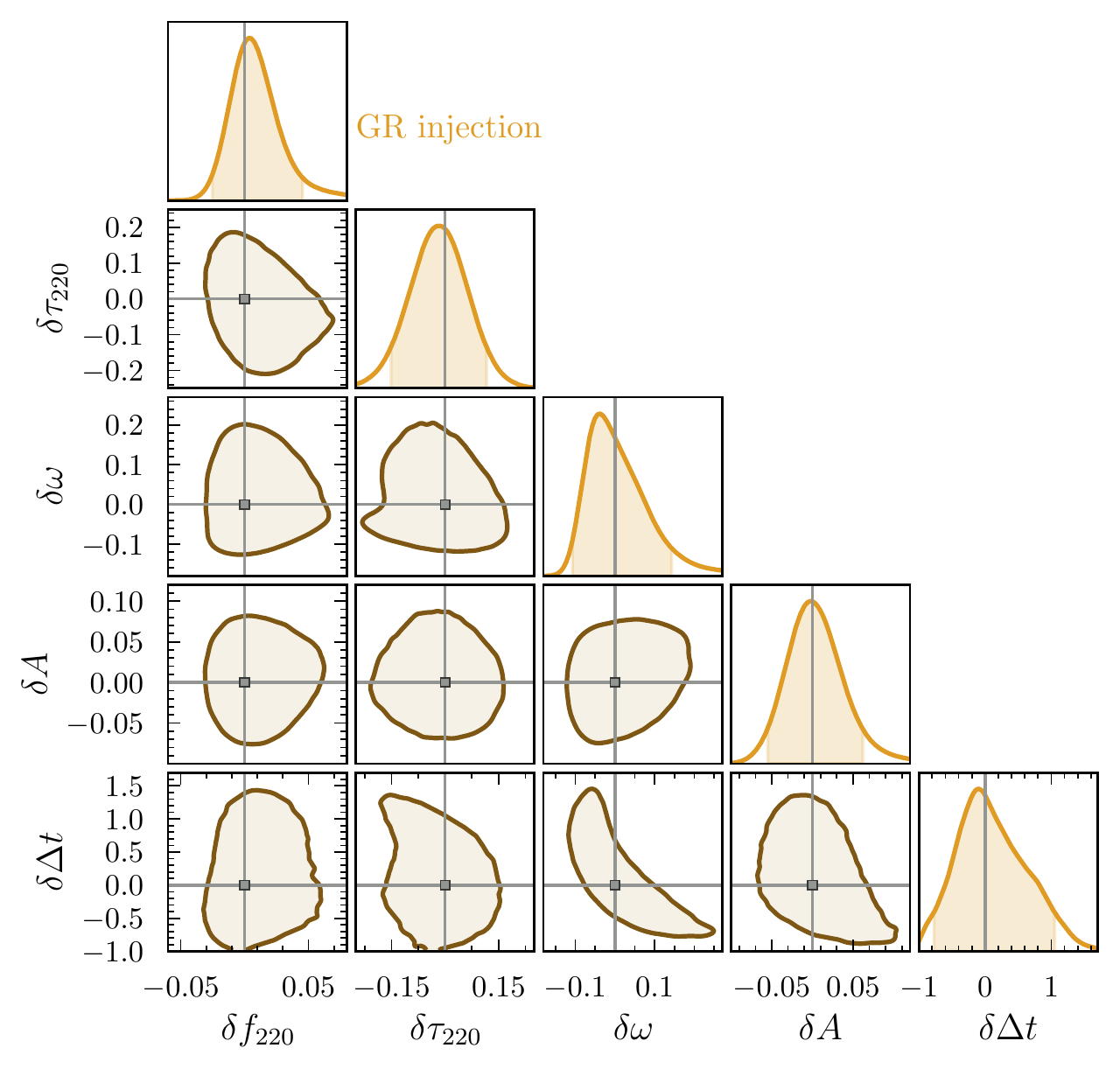}
\caption{The one- and two-dimensional posterior
distributions on the merger-ringdown parameters $\bm{\vartheta}_{\rm nGR}$.
All contours indicate 90\% credible regions.
We considered a \gr injection and recovered with the \pSEOB{} model.
%
The vertical and horizontal lines mark the injected values for the
deviation parameters, i.e.,~$\bm{\vartheta}_{\rm nGR} = 0$.
The inferred values on $\bm{\vartheta}_{\rm nGR}$ are consistent with the zero,
and their width of the marginalized posterior distribution inform us
with which accuracy we may constrain these parameters.
}
\label{fig:cornerplot_GRinj_nonGRrec}
\end{figure}

We summarize the results of the analysis in
Fig.~\ref{fig:cornerplot_GRinj_nonGRrec}, where we show the one- and
two-dimensional posterior probability distributions of the merger-ringdown
parameters $\bm{\vartheta}_{\rm nGR}$.
We find that the marginalized posterior distributions of the non-\gr parameters are
consistent with the corresponding injected values in \gr, which are indicated by the markers.
We can infer that a GW150914-like event with $\text{SNR} = 98$ would
constrain the deformation parameters in the range between 5\% (for $\delta A$ and $\delta
f_{220}$) and 20\% (for $\delta \tau_{220}$) at 90\% credible level.
In Appendix~\ref{sec:appendixA}, Fig.~\ref{fig:appendix1}, we show the
posterior distributions on the intrinsic binary parameters.

The best constrained parameter is the amplitude, $\delta A$, whereas the less
constrained parameter is the time shift, $\delta \Delta t$. For the latter,
we obtain a posterior distribution that has support onto a wide range of the prior.
This is perhaps unsurprising due to the small deviations caused by $\delta \Delta t$ in the waveform
in comparison with $\delta \omega$ (compare Figs.~\ref{fig:h22_ex_dw} and~\ref{fig:h22_ex_dt}).
We also observe a negative correlation between these two parameters and hence increasing precision on one is likely to increase uncertainty on the other; see the $\delta
\Delta t$--$\delta \omega$ panel in Fig.~\ref{fig:cornerplot_GRinj_nonGRrec}.
Together, these results suggest that considering $\delta A$ and $\delta \omega$ is
sufficient, if one is interested in doing a test of \gr only in the
plunge-merger stage of the binary's coalescence.

\subsection{Detecting deviations from general relativity}
\label{sec:results_ngr_inj_ngr_mod}

\begin{figure}[t]
\centering
\includegraphics[width=\columnwidth]{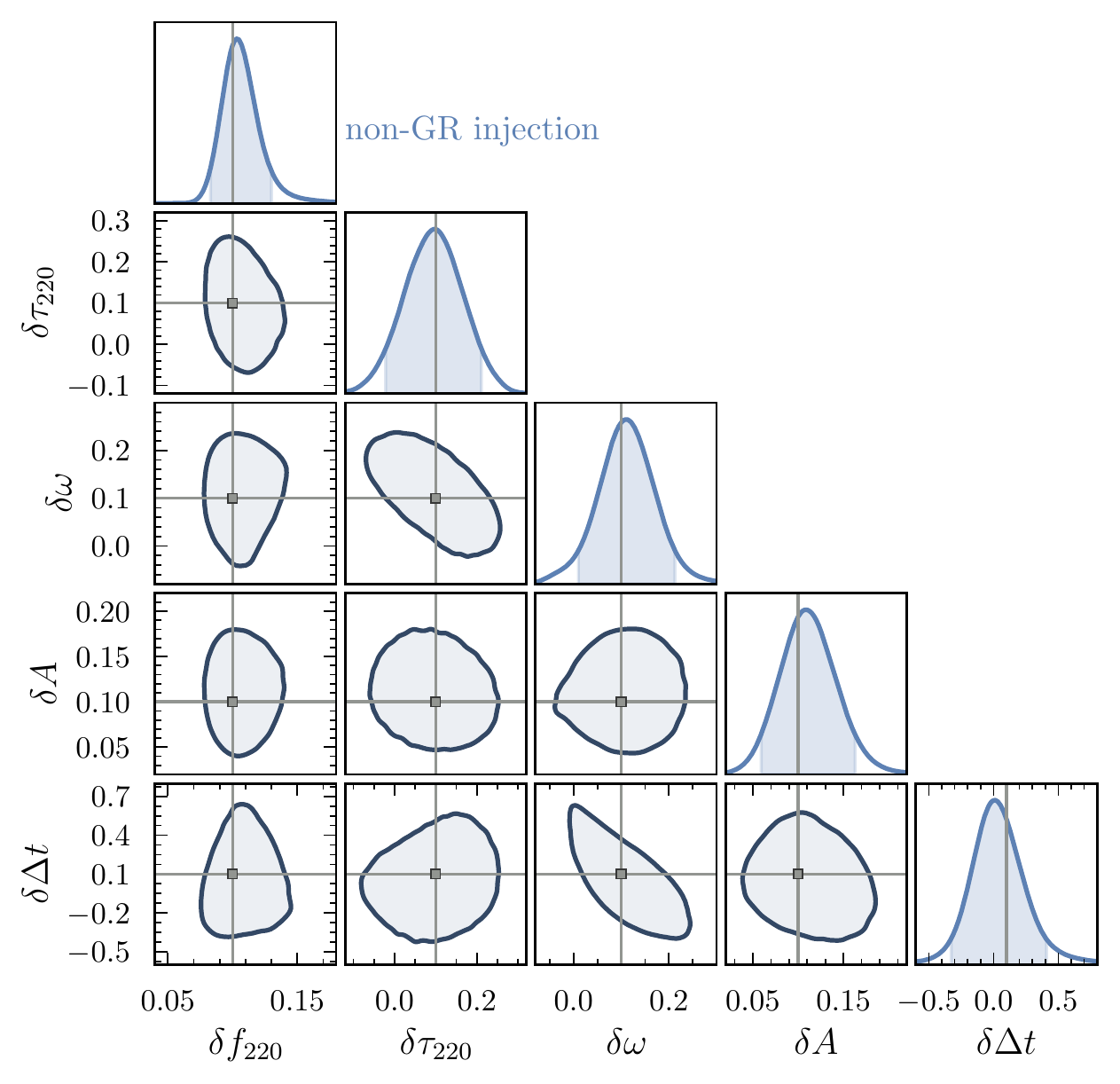}
\caption{
The one- and two-dimensional posterior distributions on the merger-ringdown parameters $\bm{\vartheta}_{\rm nGR}$. All contours correspond to 90\% credible regions.
In comparison to Fig.~\ref{fig:cornerplot_GRinj_nonGRrec}, this time we use
\pSEOB{} prepare the injection. This allow us to understand how well we
can measure the non-\gr parameters.
The vertical and horizontal lines mark the injected values for the
deviation parameters, i.e.,~$\bm{\vartheta}_{\rm nGR} = 0.1$.
The marginalized posterior distributions on $\bm{\vartheta}_{\rm nGR}$ are consistent
with their injection values.
}
\label{fig:cornerplot_10pinj_nonGRrec}
\end{figure}

We now study whether we can detect the presence of the non-GR parameters.
To do so, we inject a synthetic \gw signal where the binary parameters are shown in
Table~\ref{tab:injected_values}, $\text{SNR}=104$, and we set the merger-ringdown
parameters to be $10\%$ larger than their corresponding \gr values.

We summarize the outcome of our parameter estimation in
Fig.~\ref{fig:cornerplot_10pinj_nonGRrec}, where we show
the one- and two-dimensional posterior distributions for the
$\bm{\vartheta}_{\rm nGR}$ parameters.
We see that all posteriors are consistent with the injected values,
indicated by the markers.
Moreover, the posteriors for $\bm{\vartheta}_{\rm nGR}$ have support at their null, \gr value.
The exceptions are the amplitude $\delta A$ and the \qnm frequency $\delta f_{220}$ parameters,
which have no support at their \gr values at 90\% credible level.
This suggests that these two parameters are the most promising ones in
signaling the presence of beyond-\gr physics for GW150914-like binaries.
In fact, we will see this suggestion taking place in our analysis of GW200129 in
Sec.~\ref{sec:gw200129}.

\section{Analysis of GW150914: constraints on the plunge-merger-ringdown parameters}
\label{sec:gw150914}

Having gained some intuition on the role of the merger-ringdown parameters in
the synthetic-signal injections presented in Sec.~\ref{sec:results_synth}, we now apply the
\pSEOB~model to the analysis of real \gw events.
Our analysis, here and in Sec.~\ref{sec:gw200129}, uses the  power spectral density
of the detectors from the Gravitational Wave Open Science Center (GWOSC)~\cite{LIGOScientific:2019lzm},
and calibration envelopes as used for the analyses in Ref.~\cite{LIGOScientific:2021sio}.
We will start with GW1501914, the first \gw event observed by the LIGO-Virgo Collaboration~\cite{LIGOScientific:2016aoc}.

We will focus our analysis to two subsets of merger-ringdown parameters due
to the smaller \snr of this event (and of GW200129) in comparison to the
\snr$\approx 100$ scenarios studied in the previous section.
First, we have seen that the time-shift parameter $\delta \Delta t$ is the hardest
parameter to constrain, and that it has wide posteriors even at such large SNRs.
This motivates us to consider, among the merger parameters, only
\begin{equation}
    \bm{\vartheta}_{\rm nGR} = \{ \delta A, \, \delta \omega \}\,,
\end{equation}
to perform a \emph{``merger test} of \gr''.
Second, we performed a parameter estimation of GW150914, using
all $\bm{\vartheta}_{\rm nGR}$ parameters in Eq.~\eqref{param_non_GR}.
We found correlations between the frequency parameter $\delta \omega$ and the
\qnm deformations parameters $\delta f_{220}$ and $\delta \tau_{220}$.
Moreover, we also did a series of synthetic-signal injection studies using the binary
parameters listed in Table~\ref{tab:injected_values}, with SNR$=26$, and in Gaussian noise.
In some of these cases, we also found correlations between
$\delta \omega$ and $\delta f_{220}$ and $\delta \tau_{220}$.

In summary, these correlations arise either when the GW event has low SNR
or due to noise.
This suggest using
\begin{equation}
    \bm{\vartheta}_{\rm nGR} = \{ \delta A, \, \delta f_{220}, \, \delta \tau_{220} \} \,,
\end{equation}
to perform a \emph{``merger-ringdown test} of \gr''.

\begin{figure}[t]
\centering
\includegraphics[width=0.975\columnwidth]{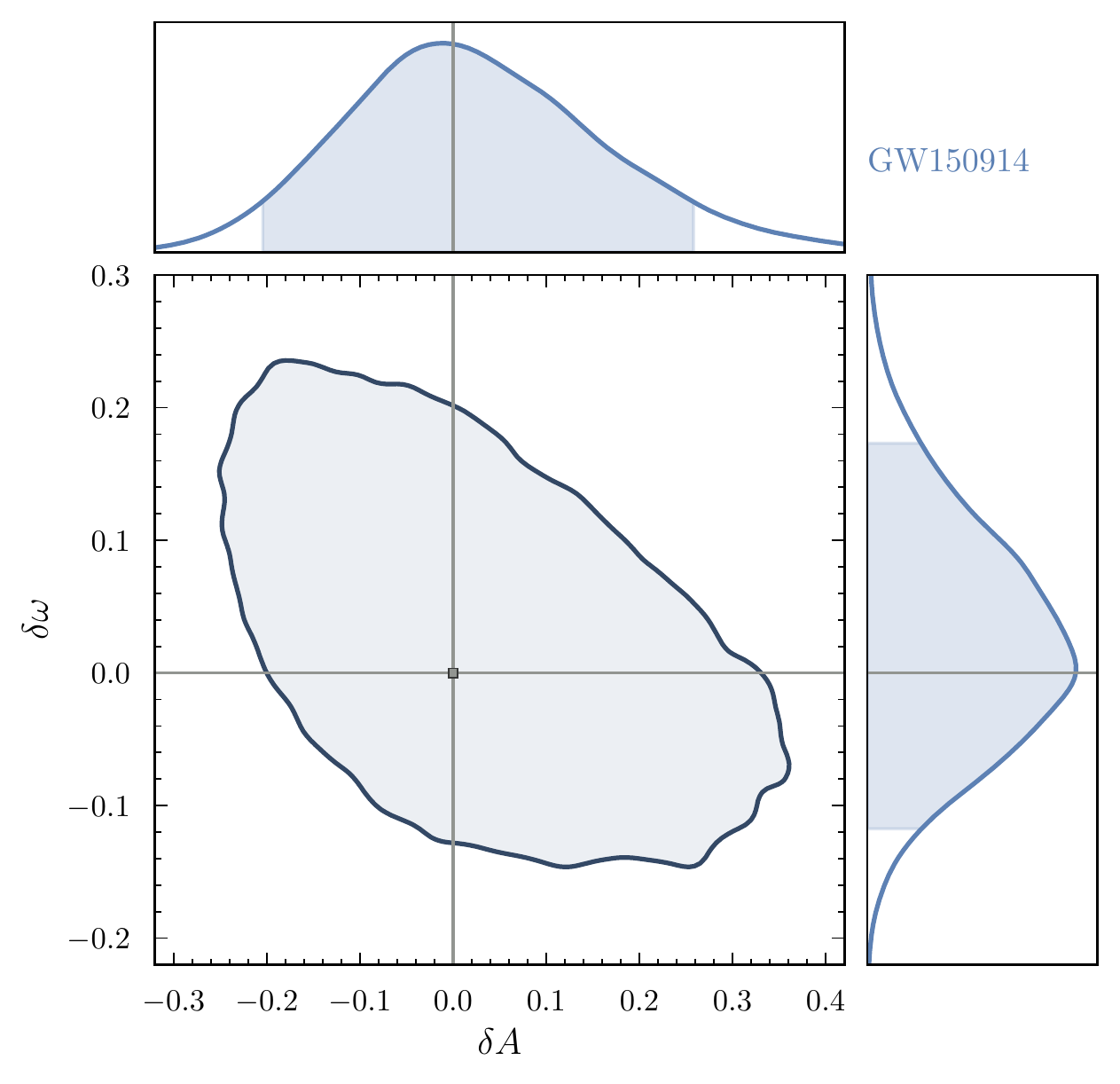}
\caption{The one- and two-dimensional posterior distributions on $\delta A$ and $\delta \omega$
obtained by analyzing GW150914. All contours correspond to 90\% credible regions.
The marginalized posterior distributions are consistent with \gr, i.e., $\delta A = \delta \omega = 0$,
identified in the plot with the horizontal and vertical lines.
We found that
$\delta A = -0.01^{+0.27}_{-0.19}$ and $\delta \omega = 0.00^{+0.17}_{-0.12}$
at 90\% credible level.
}
\label{fig:gw150914_nonGRrec_dAdw}
\end{figure}

In Fig.~\ref{fig:gw150914_nonGRrec_dAdw} we show the results of
our merger test of \gr.
The corner plot shows the one- and two-dimensional posterior probability distributions of $\delta A$
and $\delta \omega$.
The posterior distributions are consistent with the null value predicted in \gr.
We obtain from GW150914,
\begin{equation}
\delta A = -0.01^{+0.27}_{-0.19} \,, \quad  \textrm{and} \quad \delta \omega = 0.00^{+0.17}_{-0.12} \,,
\label{eq:gw150914_dAdw_results}
\end{equation}
at 90\% credible level. This shows that we \emph{already constrain deviations from \gr
around the merger time of BBH coalescences to about 20\% with present \gw events}.

\begin{figure}[t]
\centering
\includegraphics[width=\columnwidth]{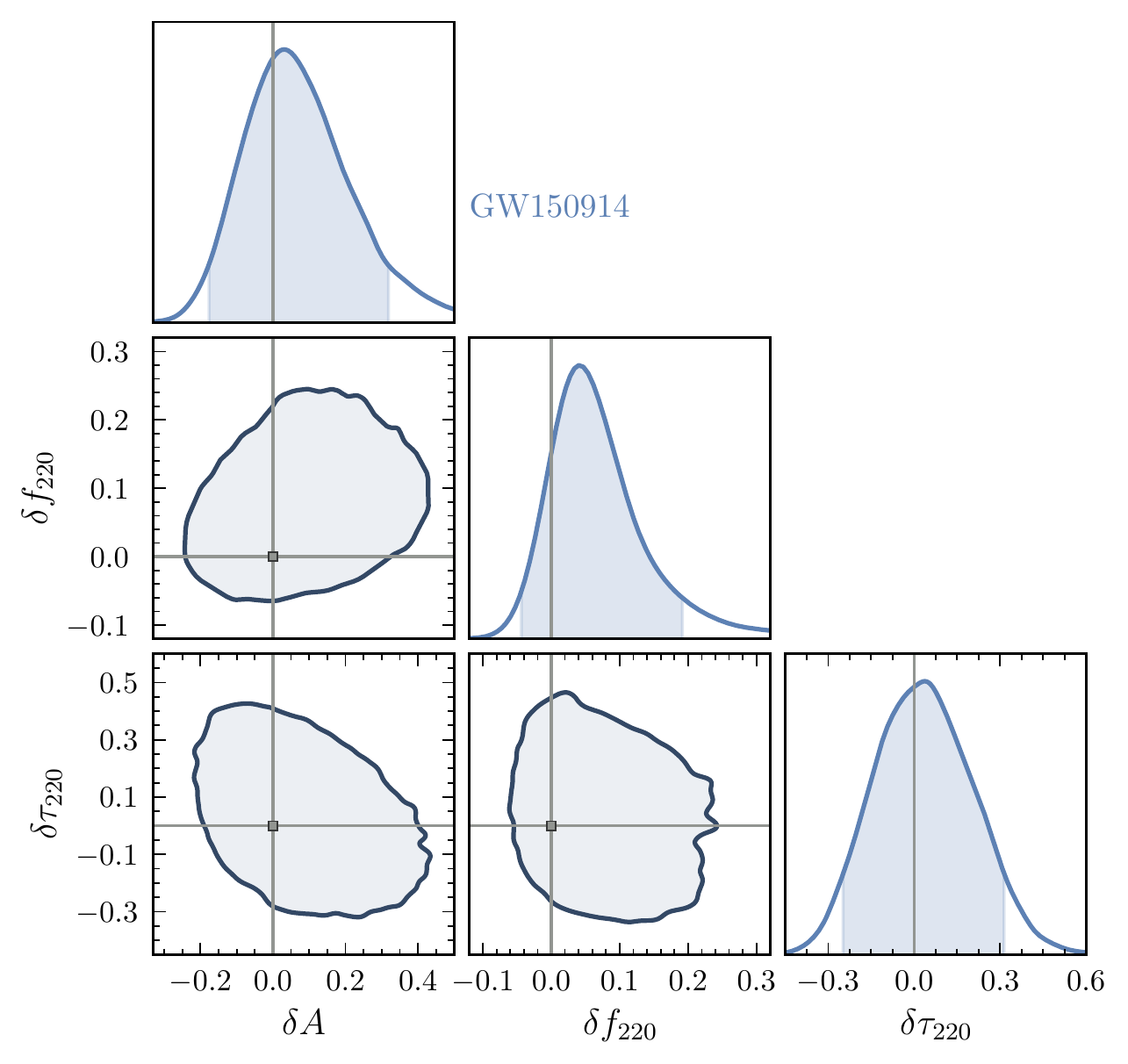}
\caption{The one- and two-dimensional posterior distributions on the merger parameter $\delta A$,
and ringdown parameters $\delta f_{220}$ and $\delta \tau_{220}$, obtained
by analyzing GW150914.
The marginalized posterior distributions are consistent with \gr, i.e., $\delta A = \delta f_{220} = \delta \omega_{220} = 0$, identified in the plot with the horizontal and vertical lines.
We found that GW150914 constrains these parameters to be
$\delta A = 0.03^{+0.29}_{-0.20}$,
$\delta f_{220} = 0.041^{+0.151}_{-0.084}$, and
$\delta \tau_{220} = 0.04^{+0.27}_{-0.29}$
at 90\% credible level.
}
\label{fig:gw150914_nonGRrec_dAdw220dt220}
\end{figure}

Figure~\ref{fig:gw150914_nonGRrec_dAdw220dt220} is a similar plot, but for
the merger-ringdown test of \gr.
Once more, we find that the inferred values of the non-\gr parameters are
consistent with \gr,
\begin{align}
\delta A = 0.03^{+0.29}_{-0.20}\,, \,
\delta f_{220} = 0.041^{+0.151}_{-0.084} \,, \,
\delta \tau_{220} = 0.04^{+0.27}_{-0.29} \,, \,
\nonumber \\
\label{eq:gw150914_dAdfdtau_results}
\end{align}
at 90\% credible level.
The bound on the amplitude parameter is similar to the one obtained in
the merger test, shown in~Eq.~\eqref{eq:gw150914_dAdw_results}. Also,
the bounds on the ringdown parameters are similar to those obtain
in Ref.~\cite{Ghosh:2021mrv} ($\delta f_{220} = 0.05^{+0.11}_{-0.07}$ and $\delta \tau_{220} = - 0.07^{+0.26}_{-0.23}$),
which had only these two quantities as its non-\gr parameters.
In Appendix~\ref{sec:appendixA}, Fig.~\ref{fig:appendix2}, we show the posterior distributions on the intrinsic binary parameters
for both tests of \gr.

When interpreting our inferences on these parameters, it is important
to note that the statistical error in our analysis ($\approx 20\%$) is larger than
the systematic error due to fitting $|h_{\lm}^{\rm NR}|$ and $\omega_{\lm}^{\rm NR}$
against \nr data, which is at most around $4\%$ with current models~\cite{Bohe:2016gbl,Cotesta:2018fcv},
depending on where one is in the $\eta$--$\chi_{\rm eff}$ parameter space.
In fact, we see that the median values of $\delta A$ and $\delta \omega$
fall within this fitting error. \emph{In conclusion, we can claim to have placed a
constraint on these non-\gr parameters with GW150914.}

\section{The case of GW200129: the importance of waveform systematics and data-quality
in tests of general relativity}
\label{sec:gw200129}

\begin{figure*}[t]
\includegraphics[width=\columnwidth]{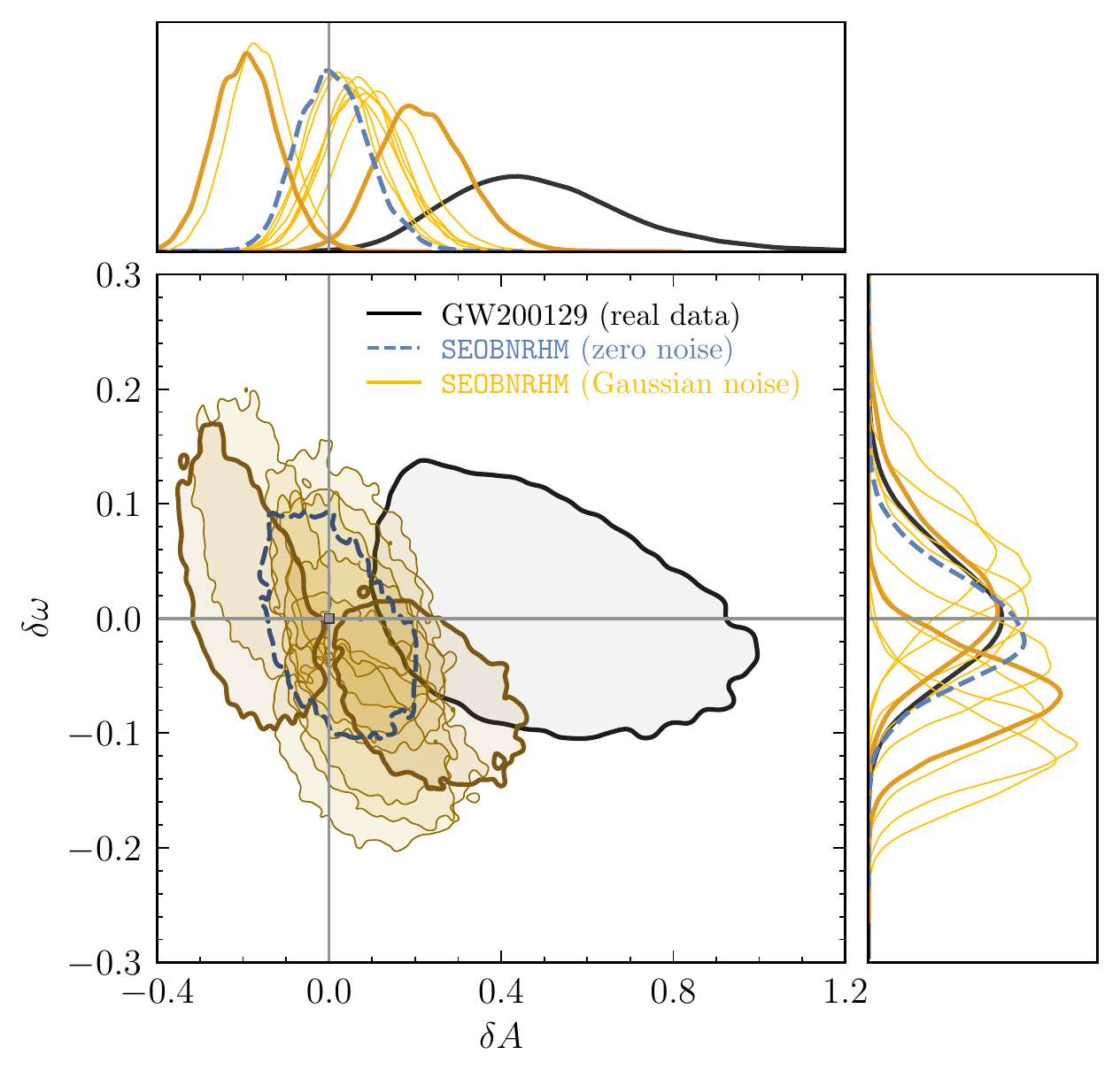}
\includegraphics[width=\columnwidth]{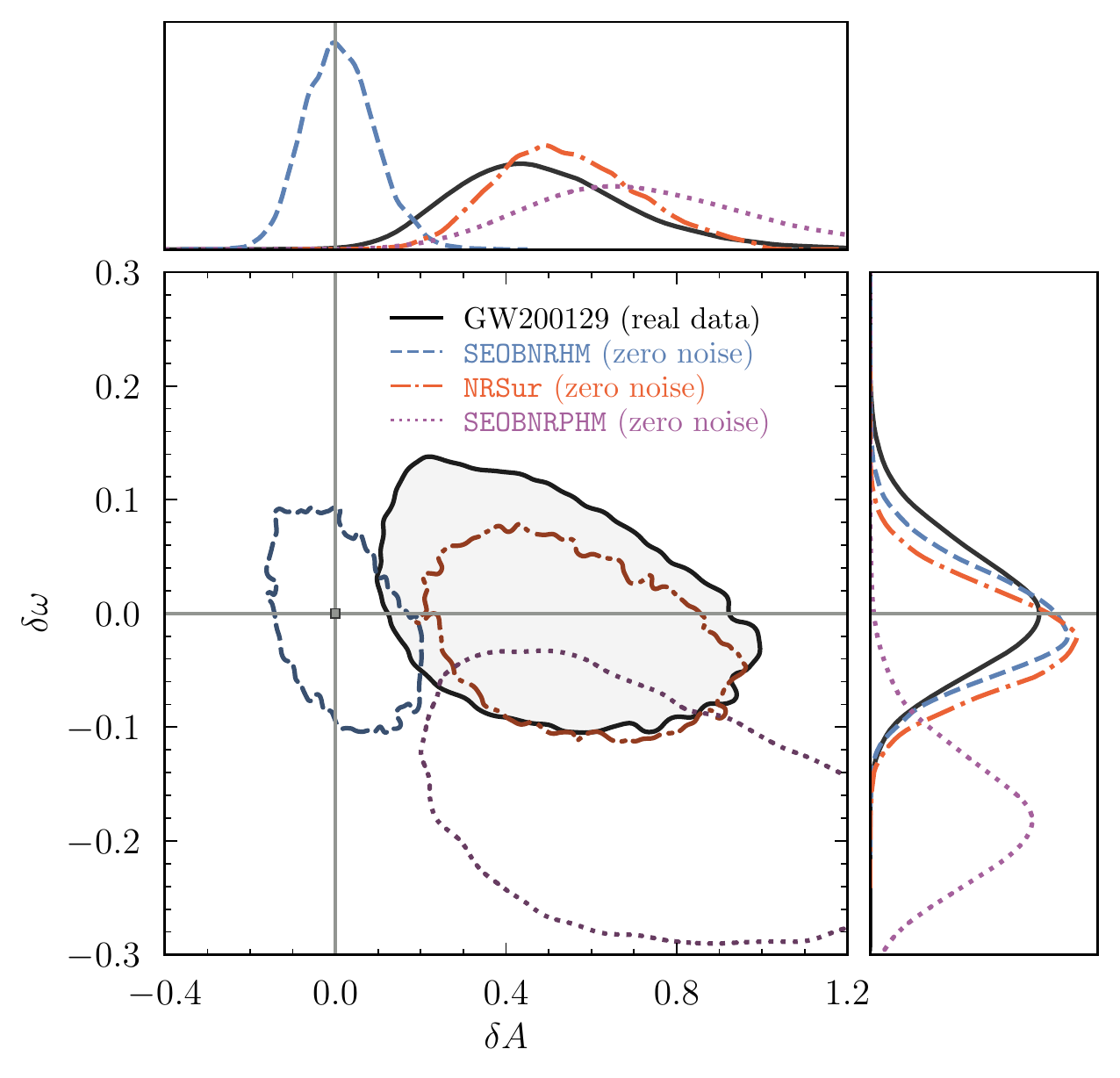}
\caption{Corner plots showing the one- and two-dimensional posterior distribution functions
for $\delta A$ and $\delta \omega$ for our studies of GW200129 and GW200129-like BBHs.
All contours indicate 90\% credible regions.
Left panel: results of our reanalysis of GW200129 data with \pSEOB{} (black
solid curves) and for a GW200129-like injection generated with \SEOB{}. For the
latter, we used the maximum-likelihood point of \LVK's original analysis of
GW200129 which employed the \IMRPHM{} model to generate the synthetic \gw
signal.
We performed the parameter estimation of these injections in zero noise (dashed
curves) and in ten Gaussian noise realizations (yellow solid curves).
Right panel: Similar, but having generated two additional GW200129-like synthetic signals with \NRSur{} (dot-dashed curves) and with \SEOBP{} (dotted curves).
Both models include spin precession effects. Observe how the posteriors distributions are in tension with \gr [marker
at $(\delta A, \, \delta \omega) = (0,\, 0)$] when we include
spin-precession effects in the synthetic data and we recover with a nonprecessing and non-\gr waveform model.
}
\label{fig:gw200129_corner_plots}
\end{figure*}

We now turn our attention to GW200129 and, following what we have learned in the previous section,
we first consider \pSEOB{} with only $\delta A$ and $\delta \omega$ as non-\gr parameters.
We show the one- and two-dimensional marginalized posteriors of these
parameters with the black solid curves in the left panel of
Fig.~\ref{fig:gw200129_corner_plots}.
We see that while our inferred value of $\delta \omega$ ($\delta \omega =
-0.002^{+0.097}_{-0.082}$ at the 90\% credible level) is consistent with \gr, our inferred
value of $\delta A$ ($\delta A = 0.44^{+0.38}_{-0.28}$ at the 90\% credible level) exhibits a
\emph{gross violation of \gr}.

Have we found a strong evidence of violation of \gr in GW200129?
Assuming that this is not the case, the apparent violation of \gr could be
either due to statistical errors or to systematic errors.
To explore the first possibility, we perform a series of synthetic-data injection studies.
As our first step, we do a parameter-estimation study in zero noise, where the
injected \gw signal is generated with \SEOB{} and we use the binary parameters
corresponding to the maximum likelihood point from the GWTC-3 data release by the
\LVK~\cite{ligo_scientific_collaboration_and_virgo_2021_5546663}
analysis of GW200129.
The \LVK analysis was done separately with two quasicircular and spin-precessing waveform models,
\SEOBP{}~\cite{Ossokine:2020kjp} and \IMRPHM{}~\cite{Pratten:2020ceb}, employing
different parameter estimation libraries, \texttt{RIFT}~\cite{Pankow:2015cra,Lange:2017wki,Wysocki:2019grj} and~\texttt{Bilby}~\cite{Ashton:2018jfp,Romero-Shaw:2020owr}, respectively.
Here, as a reference, we use the maximum likelihood point of the analysis that employed the
\IMRPHM{} model, and
we expect the results to be qualitatively similar had we used \SEOBP{}.
More specifically, because the \SEOB{} model we are using is nonprecessing, we use only the masses and
luminosity distance from the maximum-likelihood point.

The resultant posterior distributions are shown in the left panel of Fig.~\ref{fig:gw200129_corner_plots}
(dashed curves) and they are, reassuringly, consistent with \gr.
We also repeat this analysis for ten Gaussian noise realizations, using the same
synthetic \gw signal (yellow solid curves in the left panel of Fig.~\ref{fig:gw200129_corner_plots}).
Consistent with the expectations, two noise realizations yield marginalized
posteriors on $\delta \omega$ and $\delta A$ which are
not consistent with \gr at 90\% credible level
(shown by the thicker yellow solid curves).
It is worth observing how the Gaussian noise curves have qualitatively the same shapes
(spreads), with the two outliers being shifted away from $(\delta A,\, \delta
\omega) = (0,\,0)$. This is an expected behavior consistent with the stationary, Gaussian assumption of \emph{statistical noise}.
These results, hence, disfavor the possibility that the violations of \gr we are observing
are due to Gaussian noise or due to the particular binary parameters inferred for this event.
The latter alternative would have been quite unlikely in the
first place, because both GW200129 and GW150914 have similar binary parameters and SNRs,
and we have already found that GW1501914 is consistent with \gr in Sec.~\ref{sec:gw150914} (cf.~Fig.~\ref{fig:gw150914_nonGRrec_dAdw}).

As our next step, we perform two additional parameter estimation runs, in
zero noise, but now generating our synthetic \gw signal with the
\SEOBP{}~\cite{Ossokine:2020kjp} and the \NRSur{}~\cite{Varma:2019csw,scott_e_field_2022_6726994} waveform models.
Both models allow for spin precession, unlike our \pSEOB{}.
Hence, we can study if the \gr deviations we are finding are due to
systematic errors in the \gw modeling.
Once again, the maximum-likelihood point of the \LVK analysis of GW200129 using
\IMRPHM{} was used, but this time with the binary in-plane spin components included.
We show our results in the right panel of Fig.~\ref{fig:gw200129_corner_plots}.
The one- and two-dimensional posterior distributions of $\delta \omega$ and $\delta A$
are shown in dash-dotted curves for the \NRSur{} injection and with dotted curves
for the \SEOBP{} injection.
For reference, we also include the posterior distribution associated to the
\SEOB{} injection (dashed curves) and to the data from GW200129 (solid curves).
We see that these two spin-precessing \gw signals, when analyzed in zero noise,
are also in disagreement with \gr, when analyzed with our nonprecessing non-\gr model.
We also see that our results using \NRSur{} (which compares the best against \nr
simulations in its regime of validity) are in good agreement with what we obtain by analyzing the GW200129 data.
These results, compared with those obtained from the \SEOB{} injection, suggest
that \emph{the presence of spin precession in the \gr signal, biases us to find a
false evidence for beyond-\gr effects when we use a nonprecessing non-\gr model.}

Is this the full story? In Ref.~\cite{Payne:2022spz}, Payne~et~al.~revisited
the evidence of spin precession in GW200129~\cite{Hannam:2021pit}.
They concluded that the evidence for spin-precession originates from the LIGO
Livingston data, in the 20--50~Hz frequency range, alone.
This range coincides with the frequency range that displays data quality
issues, due to a glitch in the detector that overlapped in time with the signal~\cite{LIGOScientific:2021djp}.
By reanalyzing the GW200129 data with $f_{\rm low} > 50$~Hz (while leaving LIGO Hanford data intact
and not using Virgo data),
they showed that the evidence in favor of spin precession in this event disappears.
See Ref.~\cite{Payne:2022spz} for a detailed discussion.
Moreover,
a reanalysis of the LIGO Livingston glitch mitigation showed that the
difference between the spin-precessing and nonprecessing interpretations of this event is
subdominant relative to uncertainties in the glitch subtraction~\cite{Payne:2022spz}.
Since we have used the glitch-subtracted data in our parameter estimation, we are then led
to the second conclusion of our study of this event, namely that:
\emph{issues with data quality can introduce biases in non-\gr parameters,
to an extent that one can find significant false violations of \gr in \gw events detected with
present \gw observatories.}
See Ref.~\cite{Kwok:2021zny} for a recent study of this issue.

\begin{figure}[t]
\includegraphics[width=\columnwidth]{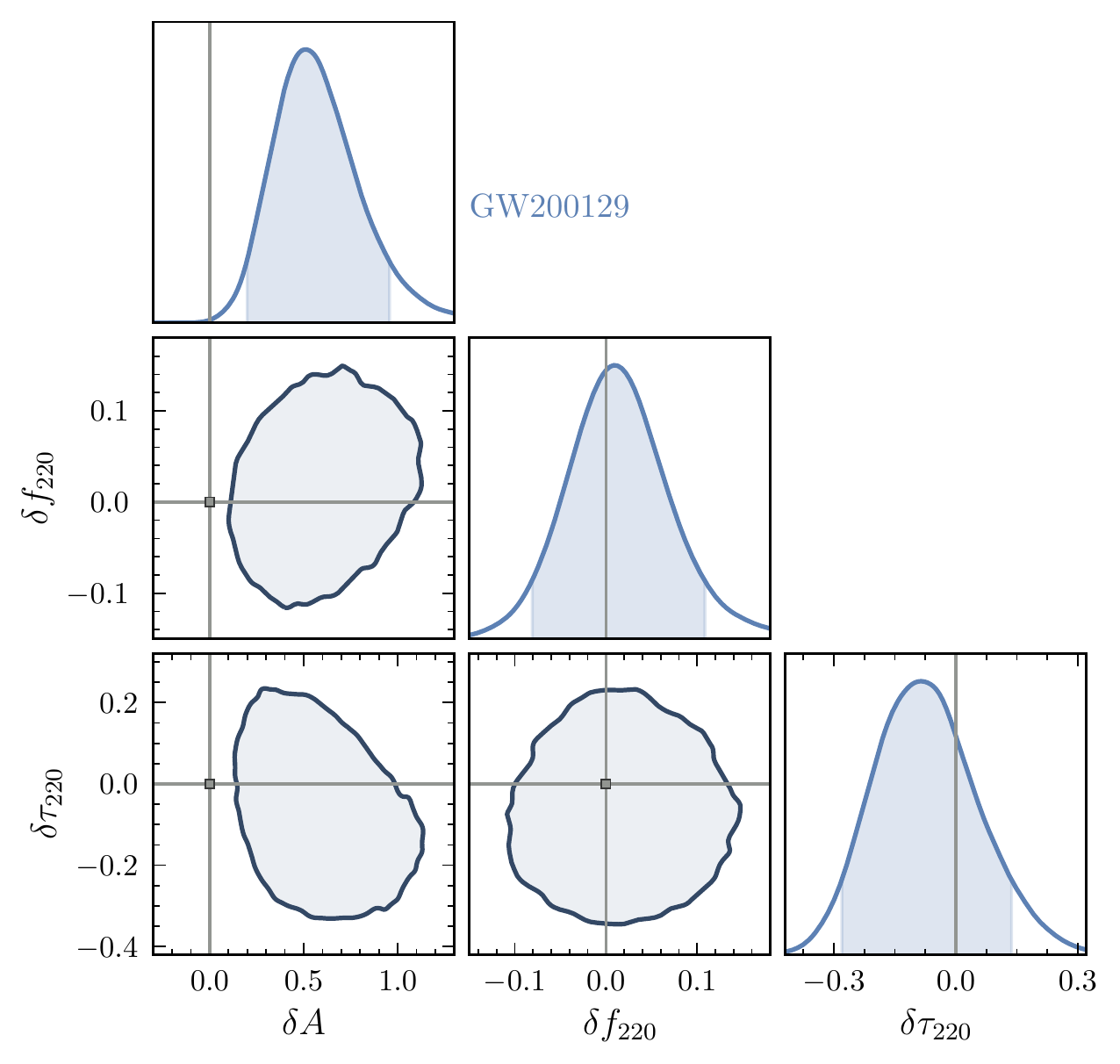}
\caption{The one- and two-dimensional posterior distribution functions
for $\delta A$, $\delta f_{220}$, and $\delta \tau_{220}$ for GW200129.
All contours indicate 90\% credible regions.
We see that while our inferred values for $\delta f_{220}$ and $\delta \tau_{220}$ are consistent with \gr,
$\delta A$ is not.
}
\label{fig:gw200129_nonGRrec_dAdw220dt220}
\end{figure}

Furthermore, we repeat here the analysis we have performed for GW150914 where
we considered $\bm{\vartheta}_{\rm nGR} = \{ \delta A, \, \delta f_{220}, \, \delta \tau_{220} \}$ as our non-\gr parameters.
For the discussion that follows, we assume that GW200129 is an
unmistakable genuine spin-precessing BBH.
We show our results in Fig.~\ref{fig:gw200129_nonGRrec_dAdw220dt220}.
We see that while our inferred values of $\delta f_{220}$ and $\delta \tau_{220}$ are consistent with
\gr at 90\% confidence level, our inference of the amplitude parameter,
$\delta A = 0.50^{+0.46}_{-0.30}$ at 90\% credible level,
remains inconsistent with \gr. Moreover, this value hardly changes from our $\{
\delta A, \, \delta \omega \}$-study, i.e.,~$\delta A = 0.44^{+0.38}_{-0.28}$, at the same credible level.

This result is interesting for two reasons. First, it indicates that the
systematic error caused by spin-precession mismodeling is robust to the
inclusion of deformations to the ringdown QNM frequencies, at least for this
event.
Second, there is a commonality between our finding for GW150914 (see Fig.~\ref{fig:gw150914_nonGRrec_dAdw220dt220})
and GW200129 (see Fig.~\ref{fig:gw200129_nonGRrec_dAdw220dt220}) namely, that in both cases
the posterior distributions of $\delta f_{220}$ and $\delta \tau_{220}$ are consistent with \gr, despite the
larger parameter space due to the inclusion of $\delta A$.
In the case in which one considers only $\delta f_{220}$ and $\delta \tau_{220}$
as non-\gr parameter, the consistency with \gr had already been established in
Ref.~\cite{Ghosh:2021mrv}, and in particular in Ref.~\cite{LIGOScientific:2021sio}; see Sec.~VIII, Fig.~14 there.\footnote{The \LVK Collaboration also does an independent analysis of the ringdown using \texttt{pyRing}.
This analysis lead to an odds ratio $\log_{10} {\cal O}^{\rm nGR}_{\rm GR} = - 0.09$ for GW200129, the
largest among all events studied~\cite{LIGOScientific:2019fpa}. A positive value would quantify the level of disagreement with \gr.}
Our analysis of these two \gw events with the new \pSEOB{} waveform model suggests the following:
\emph{the model would be able to detect deviations from nonprecessing
quasicircular \gw signals in the plunge-merger-ringdown which otherwise would not be
seen when having deformations to the ringdown only.}

We close our discussion of GW200129 with two remarks.
First, data-quality issues aside, we can think of our spin-precessing injection
studies as illustrative of what could happen in upcoming \LVK observation runs.
By doing so, we have then demonstrated the existence of a systematic error
on the non-\gr parameters caused by spin-precession mismodeling.\footnote{If the GW signal had a
smaller total-mass binary, signatures of spin precession could have been observed from the
inspiral portion of the waveform only.}
Second, although we have proposed \pSEOB{} as a means of constraining (or detecting) potential
non-\gr physics in BBH coalescences, we can also interpret the merger parameters
as indicators of our ignorance in \gr waveform modeling.\footnote{
In this interpretation, the questions we investigated
in Secs.~\ref{sec:results_ngr_inj_gr_mod} and~\ref{sec:results_gr_inj_ngr_mod} become:
\begin{enumerate*}[label={(\roman*)}]
    \item how large are the systematic errors in one's parameter inference due to \gw modeling?
    \item how large can our \gw-modeling uncertainties be such that we are still consistent
    with the ``true'' binary parameters.
\end{enumerate*}
}
More concretely, in a hypothetical scenario where \gw modelers did not know
that BBH can spin precess, an analysis of GW200129 with \pSEOB{} would
suggest that their model of the peak \gw-mode amplitudes is insufficient to
describe this event and hence be an indicative of new, nonmodeled binary
dynamics that was absent in their waveform model. They would not be able
to say that spin precession is the missing dynamics, but they would at least realize
that \emph{something} is missing.

\section{Discussions and final remarks}
\label{sec:conclusions}

We presented a time-domain \imr waveform model that accommodates parametrized deviations
from \gr in the plunge-merger-ringdown stage of nonprecessing and quasicircular BBHs.
This model generalizes the previous iterations of the \pSEOB{} model~\cite{Brito:2018rfr,Ghosh:2021mrv,Mehta:2022pcn}, which
included deviations from GR in the inspiral phase or modified the \qnm frequencies only, by introducing
deformations parameters $\bm{\vartheta}^{\rm merger}_{\rm nGR}$
that, for each GW mode, can change the time at which the \gw mode peaks, the mode frequency at this
instant, and the peak mode amplitude.
This new version of \pSEOB{} reduces to the state-of-the-art \SEOB{}
model~\cite{Bohe:2016gbl,Cotesta:2018fcv,Mihaylov:2021bpf} for nonprecessing and quasicircular
BBHs in the limit in which all deformations parameters are set to zero.

We used \pSEOB{} to perform a series of injections studies for GW150914-like events
exploring
(i) the constraints that one could place on these non-\gr parameters,
(ii) the biases introduced on the intrinsic binary parameters in case nature is
not described by \gr and we model the signal with a \gr template, and, finally,
(iii) we studied the measurability of these non-\gr parameters.

We also used \pSEOB{} in a reanalysis of GW150914 and GW200129.
For GW150914, we found that the deviations from the \gr peak amplitude and the instantaneous
\gw frequency can already be constrained to about $20\%$ at 90\% credible level.
For GW200129, we found an interesting interplay between spin precession and false violations
of \gr that manifests as a $\sim 2 \sigma$ deviation from \gr in the peak amplitude parameter.
By interpreting the evidence for spin precession in this event as due to
data-quality issues in the LIGO Livingston detector~\cite{Payne:2022spz,LIGOScientific:2021djp},
we found a further a connection between data-quality issues and false violations of \gr~\cite{Kwok:2021zny}.

These results warrant further studies on the systematic bias due to
spin precession in tests of \gr. In the context of plunge-merger-ringdown test,
this could be achieved by extending the \SEOBP{} waveform
model~\cite{Ossokine:2020kjp} to include the same set of non-\gr parameters
$\bm{\vartheta}_{\rm nGR}$ used here.
It is also natural to explore which systematic effects higher \gw modes~\cite{Pang:2018hjb} and binary eccentricity can introduce in tests of \gr.
For the latter, see Ref.~\cite{Bhat:2022amc} for work in this direction for \imr consistency tests~\cite{Hughes:2004vw,Ghosh:2016qgn} and Ref.~\cite{Saini:2022igm} in the context of deviations in the \pn \gw phasing~\cite{Yunes:2009ke,Cornish:2011ys,Agathos:2013upa}.
It would also be interesting to investigate these issues in the context of the ringdown test
within the \eob framework employed by \LVK Collaboration~\cite{LIGOScientific:2021sio}
and which relies on \pSEOB{}~\cite{Brito:2018rfr,Ghosh:2021mrv}.
This could be done by adding non-\gr deformations to the \SEOBE{} waveform model of Ref.~\cite{Ramos-Buades:2021adz}.
It would also be important to investigate whether \pSEOB{} can be used to
detect signatures of non-\gr physics, as predicted by the rapidly growing
field of \nr in modified gravity theories (see e.g.,~Refs.~\cite{Witek:2018dmd,Okounkova:2017yby,Okounkova:2019dfo,Okounkova:2019zjf,Okounkova:2020rqw,Okounkova:2022grv,East:2020hgw,Figueras:2021abd,AresteSalo:2022hua,Corman:2022xqg});
some of which predict nonperturbative departures from \gr only in late-inspiral and merger ringdown~\cite{Silva:2020omi,East:2021bqk,Doneva:2022byd,Elley:2022ept}.
One could also study what the theory-agnostic bounds we obtained with GW150914
on the amplitude and \gw frequency imply to the free parameters of various
modified gravity theories.

The deformations parameters $\bm{\vartheta}^{\rm merger}_{\rm nGR}$ in our \pSEOB{}
model should have an approximate correspondence to the phenomenological deviation parameters (from NR calibrated
values) in the ``intermediate region'' of the \texttt{IMRPhenom} waveform model used in the {\tt TIGER} pipeline~\cite{Li:2011cg,Agathos:2013upa,Meidam:2017dgf} of the
\LVK Collaboration~\cite{TheLIGOScientific:2016src,Abbott:2018lct,LIGOScientific:2019fpa,LIGOScientific:2020tif}. Such a mapping could be derived through synthetic
injection studies.
This work only introduced non-GR parameters in the EOB GW modes and only during
the plunge-merger-ringdown. Importantly, and more consistently, in the near
future we will extend the parametrization to the EOB conservative and
dissipative dynamics.

The interplay between \gw waveform systematics, characterization and
subtraction of nontransient Gaussian noises in \gw detectors, and non-\gr
physics will become increasingly important in the future.
Planned ground-based~\cite{Punturo:2010zz,Reitze:2019iox} and space-borne \gw
observatories~\cite{LISA:2017pwj} will detect \gw transients with SNRs that may
reach the thousands depending on the source.
Having all these aspects under control is a daunting task that will need to be
faced if one wants to confidently answer the question \emph{``Is Einstein still
right?''}~\cite{will2020einstein} in the stage of BBH coalescences where his
theory unveils its most outlandish aspects.

\section*{Acknowledgments}
\label{sec:acknowledgements}
%
%
We thank Héctor Estellés, Ajit Kumar Mehta, Deyan Mihaylov, Serguei Ossokine, Harald Pfeiffer,
Lorenzo Pompili, Antoni Ramos-Buades, and Helvi Witek for discussions.
We also thank Nathan Johnson-McDaniel, Juan Calder\'on Bustillo, and Gregorio Carullo for comments on this work.
%
%
We acknowledge funding from the Deutsche Forschungsgemeinschaft (DFG)~-~Project No.~386119226.
%
%
We also acknowledge the computational resources provided by the Max Planck Institute
for Gravitational Physics (Albert Einstein Institute), Potsdam, in particular,
the Hypatia cluster.
%
%
The material presented in this paper is based upon work supported by National
Science Foundation's (NSF) LIGO Laboratory, which is a major facility fully
funded by the NSF.
%
%
This research has made use of data or software obtained from the Gravitational Wave Open Science Center (\href{gwosc.org}{gwosc.org}), a service of LIGO Laboratory, the LIGO Scientific Collaboration, the Virgo Collaboration, and KAGRA. LIGO Laboratory and Advanced LIGO are funded by the United States National Science Foundation (NSF) as well as the Science and Technology Facilities Council (STFC) of the United Kingdom, the Max-Planck-Society (MPS), and the State of Niedersachsen/Germany for support of the construction of Advanced LIGO and construction and operation of the GEO600 detector. Additional support for Advanced LIGO was provided by the Australian Research Council. Virgo is funded, through the European Gravitational Observatory (EGO), by the French Centre National de Recherche Scientifique (CNRS), the Italian Istituto Nazionale di Fisica Nucleare (INFN) and the Dutch Nikhef, with contributions by institutions from Belgium, Germany, Greece, Hungary, Ireland, Japan, Monaco, Poland, Portugal, Spain. KAGRA is supported by Ministry of Education, Culture, Sports, Science and Technology (MEXT), Japan Society for the Promotion of Science (JSPS) in Japan; National Research Foundation (NRF) and Ministry of Science and ICT (MSIT) in Korea; Academia Sinica (AS) and National Science and Technology Council (NSTC) in Taiwan~\cite{LIGOScientific:2019lzm}.

\appendix

\section{Estimation of the intrinsic binary parameters}
\label{sec:appendixA}

\begin{figure}[t]
\includegraphics[width=\columnwidth]{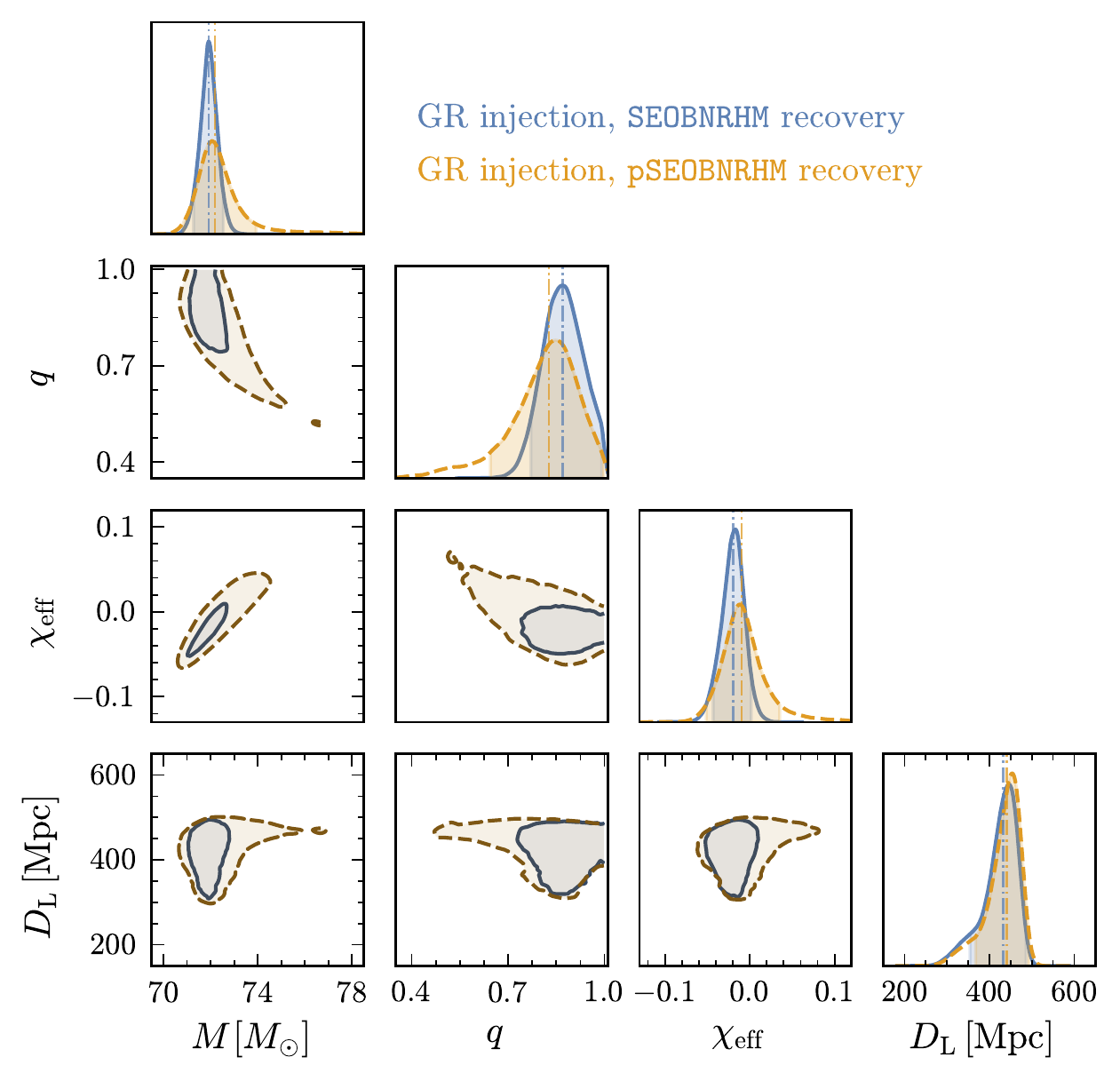}
\caption{
The one- and two-dimensional posterior distributions on the intrinsic binary
parameters of the total mas $M$, the mass ratio $q$, the effective spin $\chi_{\rm
eff}$ and the luminosity distance $D_{\rm L}$ for a \gr injection with the parameters in
Table~\ref{tab:injected_values}.
The parameter estimation is performed using \SEOB{} (solid
curves) and \pSEOB{} (dashed curves) waveform models.
All contours indicate 90\% credible regions and the vertical lines mark the
inferred median values for each parameter.
}
\label{fig:appendix1}
\end{figure}

\begin{figure}[t]
\includegraphics[width=\columnwidth]{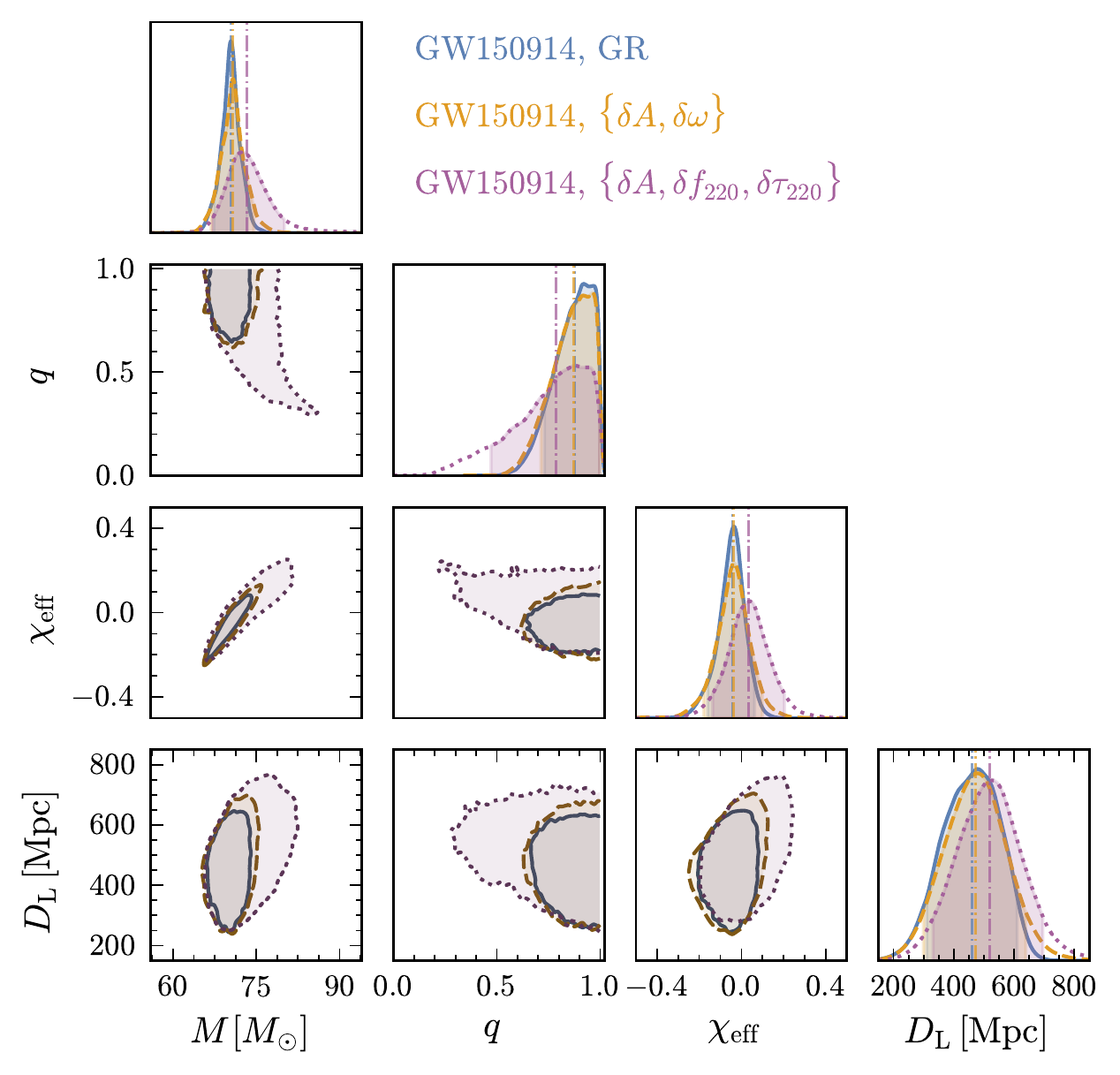}
\caption{
The one- and two-dimensional posterior distributions on the intrinsic binary
parameters of the total mass $M$, the mass ratio $q$, the effective spin $\chi_{\rm
eff}$ and the luminosity distance $D_{\rm L}$ for GW150914.
The parameter estimation is done with \SEOB~(solid curves) and \pSEOB~waveform model with
non-GR parameters $\{\delta A, \delta \omega\}$ (dashed curves) and $\{\delta A,
\delta f_{220}, \delta \tau_{220}\}$ (dotted curves).
All contours indicate 90\% credible regions and the vertical lines mark the inferred median values for each parameter.
}
\label{fig:appendix2}
\end{figure}

In this Appendix, we compare the posterior distributions on the intrinsic binary
parameters obtained using the \SEOB{} and \pSEOB{} waveform models. This complements the results shown in Sec.~\ref{sec:results_gr_inj_ngr_mod} (Fig.~\ref{fig:appendix1})
and Sec.~\ref{sec:gw150914} (Fig.~\ref{fig:appendix2}).
For simplicity, we focus on the total mass $M$, the mass ratio $q$, the effective spin
$\chi_{\rm eff}$ and the luminosity distance $D_{\rm L}$.

Figure~\ref{fig:appendix1} shows the posterior distributions on the intrinsic binary
parameters for a GR signal with the properties shown in
Table~\ref{tab:injected_values}.
The solid curves are obtained when the parameter estimation is performed with
the GR waveform model \SEOB{}, whereas the dashed curves are obtained with the
parametrized plunge-merger-ringdown waveform model \pSEOB{}
(cf.~Sec.~\ref{sec:results_gr_inj_ngr_mod}). In both cases, the SNR is 98.
We see that the 90\% confidence intervals of the posterior distributions in the
two analyses overlap in the parameter space. The most important difference is that
the 90\% credible intervals are wider in the \pSEOB{} analysis. There are also
changes to the median values of the binary parameters, as can be seen through the
vertical lines in the plot.

To be more precise, the posteriors of the \pSEOB{} model have a tail, most evidently in the
mass ratio $q$. For the mass ratio, at 90\% confidence interval, we
find $q = 0.87^{+0.12}_{-0.10}$ (for the \SEOB{} recovery) and $q = 0.84^{+0.15}_{-0.20}$ (for the \pSEOB{} recovery).
The broader posteriors, and tails, are due to the fact that the \pSEOB{} model
has five additional parameters with respect to \SEOB{}. Qualitatively,
by increasing the number of parameters in the model we increase the number of
possible waveforms that match, to some extent, the injected signal. This will
be most evidently seen in Fig.~\ref{fig:appendix2} which we discuss below.

Figure~\ref{fig:appendix2} shows the posterior distributions on the intrinsic
binary parameters for our analyses of GW150914 (cf.~Sec.~\ref{sec:gw150914}).
The solid curves are obtained when we use \SEOB{} for parameter estimation, whereas
the dashed and dotted curves are obtained when we use
\pSEOB{} with non-GR parameters $\{\delta A, \delta \omega\}$ (``merger test of GR'')
and $\{\delta A, \delta f_{220}, \delta \tau_{220}\}$ (``merger-ringdown test of GR''),
respectively.
The figure is similar to Fig.~\ref{fig:appendix1}, discussed above. Again, we
see that the 90\% confidence intervals of the posterior distributions in the
three analyses overlap in the parameter space. However, here we can see more
explicitly how the increase of extra non-GR parameters in the waveform model

\newpage

\bibliography{mrd_mods}
\end{document}